\documentclass[twocolumn,prb,superscriptaddress,showpacs]{revtex4}
\usepackage{amsmath,amssymb}
\usepackage{graphicx}
\usepackage{subfigure}
\usepackage{verbatim}
\usepackage{amsfonts}
\usepackage{dcolumn}
\usepackage{bm}
\usepackage{color}
\usepackage[colorlinks,citecolor=blue]{hyperref}

\begin{document}

\title{\large \bf Symmetry Protected Topological Phases in Spin Ladders with Two-body Interactions}

\author{Zheng-Xin Liu}
\affiliation{Institute for Advanced Study, Tsinghua University, Beijing, 100084, China}
\affiliation{Department of Physics, Massachusetts Institute of Technology, Cambridge, Massachusetts 02139, USA}

\author{Zhen-Biao Yang}
\email{zbyang@ustc.edu.cn}
\affiliation{Key Laboratory of Quantum Information, University of Science and Technology of China,
CAS, Hefei, Anhui, 230026, People's Republic of China}

\author{Yong-Jian Han}
\affiliation{Key Laboratory of Quantum Information, University of Science and Technology of China,
CAS, Hefei, Anhui, 230026, People's Republic of China}

\author{Wei Yi}
\email{wyiz@ustc.edu.cn}
\affiliation{Key Laboratory of Quantum Information, University of Science and Technology of China,
CAS, Hefei, Anhui, 230026, People's Republic of China}

\author{Xiao-Gang Wen}
\email{wen@dao.mit.edu}
\affiliation{Department of Physics, Massachusetts Institute of Technology, Cambridge, Massachusetts 02139, USA}
\affiliation{Perimeter Institute for Theoretical Physics, Waterloo, Ontario, Canada N2L 2Y5}

\begin{abstract}
Spin-1/2 two-legged ladders respecting inter-leg exchange symmetry $\sigma$ and
spin rotation symmetry $D_2$ have new symmetry protected topological (SPT)
phases which are different from the Haldane phase.  Three of the new SPT phases
are $t_x,t_y,t_z$, which all have symmetry protected two-fold degenerate edge
states on each end of an open chain.  However, the edge states in different
phases have different response to magnetic field.  For example, the edge states
in the $t_z$ phase will be split by the magnetic field along the $z$-direction,
but not by the fields in the $x$- and $y$-directions.  We give the Hamiltonian
that realizes each SPT phase and demonstrate a proof-of-principle quantum
simulation scheme for Hamiltonians of the $t_0$ and $t_z$ phases based on the
coupled-QED-cavity ladder.

\end{abstract}


\maketitle

\section{Introduction}
Symmetry protected topological (SPT) phases are formed by gapped short-range-entangled
quantum states that do not break any symmetry.~\cite{GuWen09} Contrary to the trivial case, quantum
states in the non-trivial SPT phases cannot be transformed into direct
product states via local unitary transformations which commute with
the symmetry group. Meanwhile, two different states belong to the same SPT
phase if and only if they can be transformed into each other by symmetric local
unitary transformations.~\cite{ChenGuWen10} A nontrivial SPT phase is different
from the trivial SPT phase because of the existence of non-trivial edge states on
open boundaries. This non-trivial property is protected by symmetry, because
once the symmetry is removed, the SPT phases can be smoothly connected to the
trivial phase without phase transitions.~\cite{pollmann} The well known Haldane
phase \cite{Haldanephase} is an example of SPT phase in 1-dimension (1D).
Topological insulators \cite{KM0501,BZ,KM0502,MB0706,FKM0703,QHZ0837} are examples of SPT
phases in higher dimensions.

The bosonic SPT phases are classified by projective representations (which describe the edge states at open boundaries or at the positions of impurities \cite{Hagiwara-1990,WhiteHuse1993,LechOrig}) of the
symmetry group in 1D,~\cite{CGW} and by group cohomology theory in higher
dimensions.~\cite{CGLW} We also have a systematic understanding of free fermion
SPT phases \cite{TIclass} and some interacting fermion SPT phases.~\cite{Kitaev,GW,Evelyn,Fermion2D} Using those general results, fourteen new 1D
SPT phases protected by $D_2$ spin rotation and time reversal symmetry are
proposed in Ref.~\onlinecite{LCW}.

In this work, we will discuss two-legged spin-1/2 ladder models with two-body
anisotropic Heisenberg interactions respecting $D_2\times\sigma$ symmetry. Here $\sigma$ is the
inter-chain exchanging symmetry and $D_2=\{E, R_x, R_y,
R_z\}$, where $E$ is the identity, $R_x$ ($R_y, R_z$) is a $180^\circ$ rotation
of the spin along $x$ ($y, z$) direction. The symmetry $D_2\times \sigma$
protects seven non-trivial SPT phases. Four of them, $t_0$, $t_x$, $t_y$, $t_z$,
can be realized in spin-1/2 ladder models. The $t_0$ phase is the Haldane
phase, and the $t_x$, $t_y$, $t_z$ phases are new
because of their different edge states. We provide a simple two-body
Hamiltonian for each SPT phase, and study the phase transitions between these
phases. We also discuss possible physical realizations of the SPT
Hamiltonians and demonstrate a proof-of-principle implementation scheme based
on coupled quantum electrodynamics (QED) cavity ladder.

The paper is organized as follows: in Sec. II, we introduce the projective representations of the underlying symmetry
group of the two-legged spin-1/2 ladder and discuss the possible SPT phases. We then numerically study the phase diagram and the phase transitions
of the spin-ladder Hamiltonians in Sec. III. In Sec. IV, we discuss the physical realizations of the SPT Hamiltonians and demonstrate a proof-of-principle quantum simulation scheme based on QED cavity ladder. Finally, we summarize in Sec. V.

\section{Projective representations of the symmetry group and SPT phases}
The symmetry group $D_2\times\sigma$ for our two-legged spin-1/2 ladder is
Abelian. All its eight representations are 1-dimensional.
The following two-spin states on a rung form four different 1D
representations of the symmetry group: $|0,0\rangle={1\over\sqrt2}(|\uparrow_1
\downarrow_2\rangle-|\downarrow_1 \uparrow_2\rangle)$, $|1,x\rangle
={1\over\sqrt2}(|\downarrow_1 \downarrow_2\rangle-|\uparrow_1
\uparrow_2\rangle)$, $|1,y\rangle={i\over\sqrt2} (|\downarrow_1\downarrow_2 \rangle
+|\uparrow_1 \uparrow_2\rangle)$, and $|1,z\rangle={1\over\sqrt2}
(|\uparrow_1 \downarrow_2\rangle+ |\downarrow_1 \uparrow_2\rangle)$. The
subscripts $1,2$ label the different spins on the same rung.  The group
$D_2\times\sigma$ has eight projective representations (see Tab.~\ref
{tab:UnitprojD2h}), which describe eight different SPT phases \cite{note2} of
spin ladder models.~\cite{CGW} One of the projective representations is
one-dimensional and trivial. The other seven non-trivial ones are two-dimensional,
which describe the seven kinds of two-fold degenerate edge states of the seven
non-trivial SPT phases.

\begin{table}[t]
\caption{Projective representations and the corresponding SPT phases for $D_2\times \sigma$ group. The active operators can split the degeneracy of the ground states. The operator $O_\pm=O_1\pm O_2$, where 1,2 are the labels of the two spins at a rung, and $SS_-$ means $\mathbf S_{1,i}\cdot\mathbf S_{1,i+1}-\mathbf S_{2,i}\cdot\mathbf S_{2,i+1}$.}
\label{tab:UnitprojD2h}
\begin{ruledtabular}
\begin{tabular}{c|ccc|c|c}
&$R_z$&$R_x$&$\sigma$&active operators&SPT phases\\
\hline\hline
$E_0 $       &     1      &      1     &      1      & &rung-singlet\footnotemark, $t_x\times t_x$, ...\\
$E_1 $       &     I      & $i\sigma_z$& $ \sigma_y$&$(S^{z}_{-},S^{z}_{+},   SS_-)$&$t_x\times t_y$\\ 
$E_2 $       & $ \sigma_z$&     I      & $i\sigma_y$&$(S^{x}_{-},S^{x}_{+},   SS_-)$&$t_y\times t_z$\\ 
$E_3 $       & $i\sigma_z$& $ \sigma_x$&     I      &$(S^{x}_{+},S^{y}_{+},S^{z}_{+})$&$t_0$, $t_x\times t_y\times t_z$ \\
$E_4 $       & $ \sigma_z$& $i\sigma_z$& $i\sigma_x$&$(S^{y}_{+},S^{y}_{-},   SS_-)$& $t_x\times t_z$\\ 
$E_5 $       & $i\sigma_z$& $ \sigma_x$& $i\sigma_x$&$(S^{x}_{+},S^{y}_{-},S^{z}_{-})$&$t_x$\\
$E_6 $    & $i\sigma_z$& $i\sigma_x$& $ \sigma_z$&$(S^{x}_{-},S^{y}_{-},S^{z}_{+})$&$t_z$\\
$E_7 $    & $i\sigma_z$& $i\sigma_x$& $i\sigma_y$&$(S^{x}_{-},S^{y}_{+},S^{z}_{-})$&$t_y$
\footnotetext{If the system has translational symmetry, there are four different rung-singlet phases: rung-$|0,0\rangle$ phase, rung-$|1,x\rangle$ phase, rung-$|1,y\rangle$ phase and rung-$|1,z\rangle$. If the system does not have translational symmetry, then there is no difference between the four rung-singlet phases and there will be only one rung-singlet phase.}
\end{tabular}
\end{ruledtabular}
\end{table}


The degeneracy at the edge in each non-trivial SPT phase is protected by
the symmetry.  To lift such a degeneracy, we need to add perturbations to break
the $D_2\times\sigma$ symmetry. The operators that split the edge degeneracy
are called active operators.~\cite{LCW} The active operators are different in different SPT phases, which provide
experimental methods to probe different SPT phases.


We will mainly study five of the eight SPT phases which can be realized in
two-legged spin-1/2 ladders: one trivial SPT phase (corresponding to $E_0$) and four nontrivial ones $t_0$, $t_x$, $t_y$, $t_z$ (corresponding to $E_3, E_5,E_6,E_7$ respectively).
The $t_0$ phase is equivalent to the spin-1 Haldane phase, where the edge
states behave like a spin-1/2 spin and will be polarized by a magnetic field in
arbitrary direction.  However, in the $t_z$ phase, the edge states will
only respond to the $z$ component of a {homogeneous} magnetic field
(`homogeneous' means that the field takes the same value at the two sites of a
rung). That is to say, the ground state degeneracy will be lifted by $B_z$ but
not by $B_x$ or $B_y$ (see Appendix A).  The $t_x$ and $t_y$ phases are defined similarly. From
the response of the edge states to external magnetic field, we can distinguish
the four phases $t_0, t_x, t_y, t_z$.~\cite{note1}

The other three SPT phases can be realized by stacking two of the above ones: the SPT phases corresponding to $E_1$, $E_2$, $E_4$ can be realized by stacking $t_x$ and $t_y$, $t_y$ and $t_z$, $t_x$ and $t_z$, respectively. The $t_0$ phase can also be realized by stacking three phases $t_x,t_y,t_z$.


The nontrivial SPT phases $t_0,t_x,t_y,t_z$ can be realized in two-legged spin-1/2 ladders.
The Hamiltonian that realize these phases is simply given as $H=H_{L}+H_{T}$, where $H_{L}$ is the interaction along the leg (or longitudinal direction) and $H_{T}$ is the interaction along the rung (or
transverse direction). We assume $H_L$ takes the following form,
\begin{eqnarray*}\label{h}
H_{L}&=&J\sum_{i}\left[(\delta S_{1,i}^x S_{1,i+1}^x + S_{1,i}^y S_{1,i+1}^y + S_{1,i}^z S_{1,i+1}^z)\right.\nonumber\\
&&\left.\ \ \ \ \ \ + (\delta S_{2,i}^xS_{2,i+1}^x + S_{2,i}^y S_{2,i+1}^y+S_{2,i}^z S_{2,i+1}^z)\right],
\end{eqnarray*}
where $J>0$, $\delta\sim1$, and $S^m_{j,i}$ ($m=x,y,z$) is the spin operator at the $j$th leg and $i$th rung.

By tuning the interaction $H_{T}$, we can obtain four SPT phases $t_0, t_x, t_y, t_z$. For example, the following model
\begin{eqnarray}
H_0 &=& H_{L} +\lambda\sum_i\mathbf S_{1,i}\cdot\mathbf S_{2,i}.\label{t0H}
\end{eqnarray}
can realize the $t_0$ phase.  When $\lambda>0$, the rung-singlet state
($|0,0\rangle$) is lower in energy and the system falls into the rung-$|0,0\rangle$
phase, which is a trivial SPT phase. When $\lambda<0$, the rung-triplet
($|1,x\rangle, |1,y\rangle, |1,z\rangle$) are lower in energy, and effectively
we obtain a spin-1 anisotropic Heisenberg model, which belongs to the Haldane phase $t_0$.

By partially flipping the sign of interactions along the rung, we obtain the
Hamiltonian for the $t_z$ phase:
\begin{eqnarray}
H_z&=&H_{L} + \lambda\sum_i(-S_{1,i}^xS_{2,i}^x-S_{1,i}^yS_{2,i}^y+S_{1,i}^zS_{2,i}^z).\label{Hz}
\end{eqnarray}
When $\lambda>0$, it falls into a trivial SPT phase which corresponds to the
rung $|1,z\rangle$ product state. When $\lambda<0$, the low energy degrees of
freedom in each rung are given by three states $|1,x\rangle, |1,y\rangle,
|0,0\rangle$, and the resultant model belongs to the $t_z$ phase. Note that the Hamiltonians (\ref{t0H}) and (\ref{Hz}) (with the respective ground states $t_0$ and $t_z$) can be transformed into each other by the unitary transformation: $U_1(\pi)=\exp\left(i\pi\sum_iS^z_{1,i}\right)$. However, this does not mean that $t_0$ and $t_z$ belong to the same phase since $U_1(\pi)$ does not commute with the symmetry group $D_2\times \sigma$. Applying similar arguments, we may have $H_x=H_{L} + \lambda\sum_i(S_{1,i}^xS_{2,i}^x-
S_{1,i}^yS_{2,i}^y -S_{1,i}^zS_{2,i}^z)$ for the $t_x$ phase, and $H_y=H_{L} +
\lambda\sum_i(-S_{1,i}^xS_{2,i}^x+ S_{1,i}^yS_{2,i}^y-S_{1,i}^zS_{2,i}^z)$ for
the $t_y$ phase. Tab. \ref{tab:UnitprojD2h} shows that different SPT phases
have different active operators. For example, in the $t_0$ phase, the edge
degeneracy can be lifted by either $S^x_+$ or $S^y_+$ or $S^z_+$. This means
that the edge states can be polarized by a homogeneous magnetic field along any
direction. In the $t_z$ phase, $S^x_+$ and $S^y_+$ are not active operators,
indicating that a weak homogeneous magnetic field in the $x$-$y$ plane will not
split the edge degeneracy.  These properties can be verified by a finite-size
exact diagonalization study of the Hamiltonian (\ref{t0H}) and (\ref{Hz}).~\cite{LCW}

\section{phase transitions and the phase diagram}

To study the phase transitions, we consider the following
model which contains both $t_0$ and $t_z$ phases,
\begin{eqnarray}\label{lambda}
H=H_{L}+\sum_i[\lambda_{xy}(S_{1,i}^xS_{2,i}^x+S_{1,i}^yS_{2,i}^y)+\lambda_z S_{1,i}^zS_{2,i}^z].
\end{eqnarray}
Since the $\lambda_{xy}$-$\lambda_z$ plane phase diagrams of above model are quite different for $\delta\neq1$ and $\delta=1$, we will discuss them separately.

\subsection{the case $\delta\neq1$}

\subsubsection{The phase daigram}
\begin{figure}[t]
\centering
\includegraphics[width=3.4in]{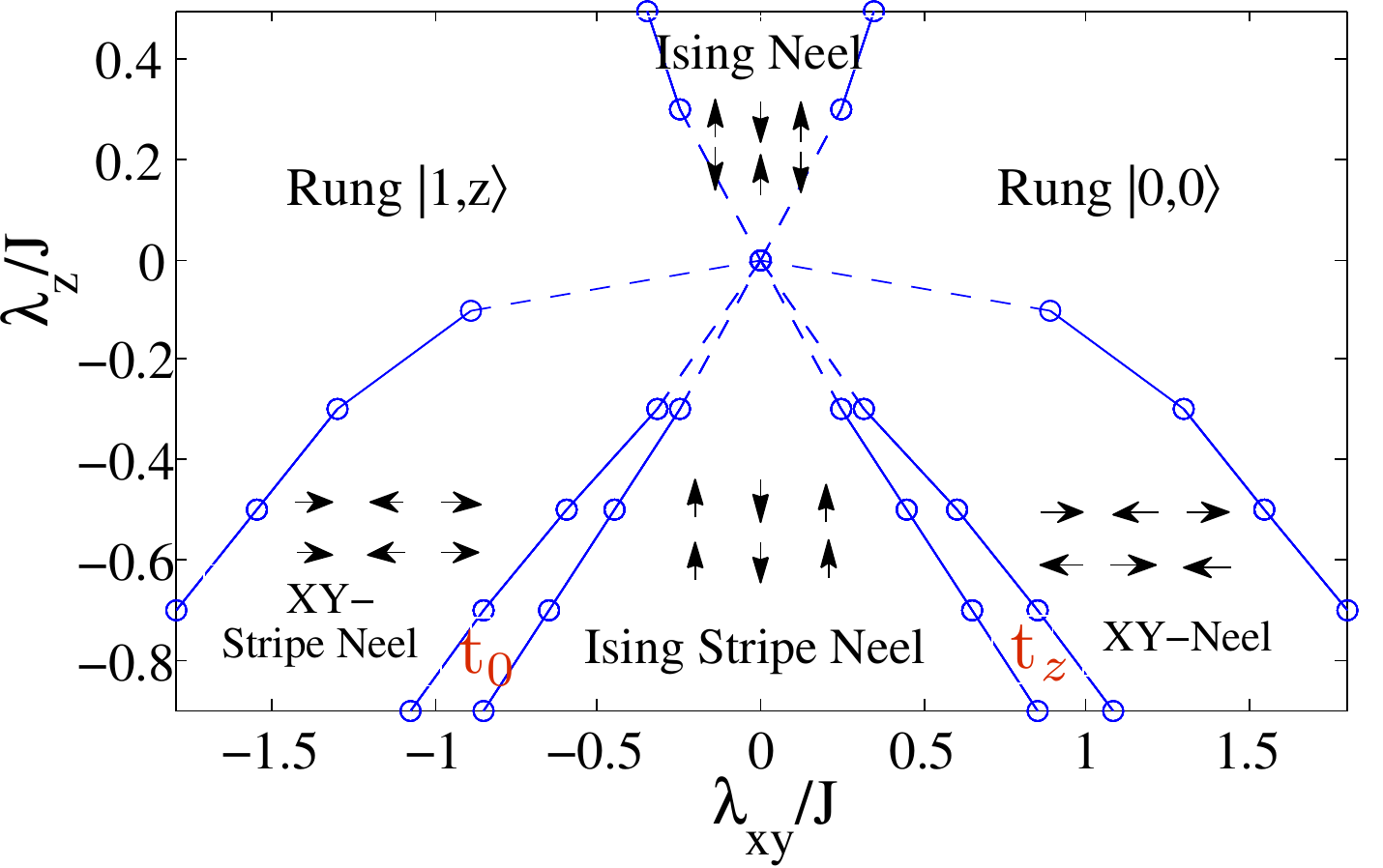}
\caption{(Color online) The phase diagram of the $\lambda_z$-$\lambda_{xy}$
plane with $\delta=0.9$. The point $(\lambda_{xy},\lambda_z)=(0,0)$ is a multi-critical point.
Near this point, the phase boundaries are hard to determine. The dashed line
means that the phase boundaries are not accurate.
The two `XY' phases are gapped and magnetically ordered in $y$-direction because $\delta\neq1$ introduces anisotropy in the $x$-$y$ plane. The $t_0$/$t_z$ SPT phase locates between an Ising-type ordered phase and an XY-type ordered phase.
}
\label{fig:phasedg}
\end{figure}

We only consider the case that anisotropy is weak, so we set $\delta=0.9$ for simplicity. The phase diagram Fig.~\ref{fig:phasedg} (with $\delta=0.9$) is obtained via infinite time-evolving
block decimation (iTEBD) algorithm.~\cite{ITEBD} The phase diagram is symmetric along the line $\lambda_{xy}=0$, because the model with $-\lambda_{xy}$ can be obtained from the one with $\lambda_{xy}$ by the unitary transformation $U_1(\pi)$.
The origin $(\lambda_{xy},\lambda_z)=(0,0)$ is a multi-critical point linking all the
phases.  On the upper half plane $\lambda_z>0$, there are only three phases.
The lower half plane is more interesting. The limit $\lambda_z\to-\infty$
corresponds to the Ising Stripe-Neel phase, while $\lambda_{xy}\to\pm\infty$
corresponds to the rung-$|0,0\rangle$/rung-$|1,z\rangle$ phase.

In the intermediate region, we have two XY-like phases and two SPT phases: XY-stripe Neel phase and XY-Neel phase.
For $\delta\neq1$, the exchange interactions along the legs are anisotropic in the $x$-$y$ plane, consequently the
two XY-like phases have a finite excitation gap and are ordered in $y$-direction if $\delta<1$ (or ordered in $x$-direction if $\delta>1$). The $t_0$/$t_z$ phase is located between the XY-stripe Neel/XY-Neel phase and the Ising stripe Neel phase.
The phase diagram at $\lambda_z=-0.5J$ is also shown in Fig.~\ref{fig:JxJz}, where we illustrate the symmetry breaking orders and the entanglement spectrum ($\Delta\rho=\rho_1-\rho_2$ is the difference
between the two biggest Schmidt eigenvalues) in each phase. For $\delta=1$, the two $XY$ phases vanish, and the SPT phase $t_0/t_z$ touches the trivial phase rung-$|1,z\rangle$/rung-$|0,0\rangle$ directly (we will study this case in detail later).

\subsubsection{Semiclassical explanation of the phase diagram}

Notice that both of the SPT phases are sandwiched by two ordered phases. This
suggests that they originate from quantum fluctuations caused by the
competition between the different classical orders. To understand this point
better, it will be interesting to compare the phase diagram
Fig.~\ref{fig:phasedg} with that from the semiclassical approach in which the ground
state is approximated by a direct product state. A semiclassical phase diagram
(Fig.~\ref{fig:class} with $\delta=0.9,\ \lambda_z=-0.5J$) is obtained by minimizing the
energy of the following trial wave function,
\begin{eqnarray*}
|\psi\rangle_{\mathrm{sc}}&=&\prod_i(a_1\phi_{1,2i}+a_2\phi_{2,2i}+a_3\phi_{3,2i}+a_4\phi_{4,2i})\nonumber\\
&&\otimes(a_2\phi_{1,2i+1}+a_1\phi_{2,2i+1}-a_3\phi_{3,2i+1}+a_4\phi_{4,2i+1}),
\end{eqnarray*}
where $\phi_1=-{1\over\sqrt2}(|1,x\rangle+i|1,y\rangle)$, $\phi_2={1\over\sqrt2}(|1,x\rangle-i|1,y\rangle)$, $\phi_3=|1,z\rangle$, $\phi_4=|0,0\rangle$ are four bases of each rung, and $a_1, a_2, a_3, a_4$ are four trial parameters.
\begin{figure}[t]
  \centering
    \includegraphics[width=3.4in]{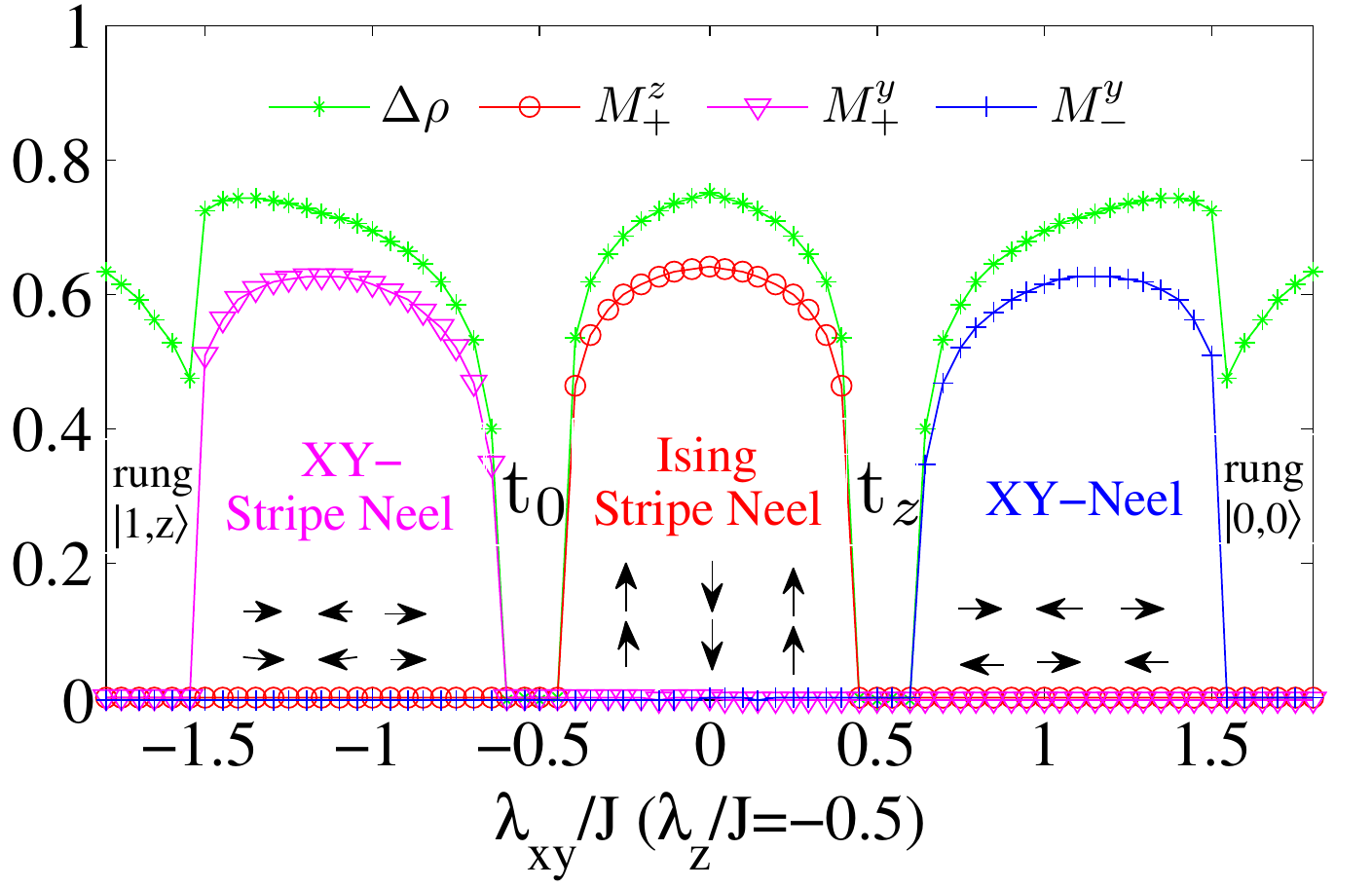}
\caption{(Color online) Phase transitions with $\delta=0.9,\ \lambda_z=-0.5J$.
The order parameters are defined as $M_\pm^m=|\langle S_1^m\pm S_2^m\rangle|$.
The green line with asterisks shows the information of entanglement spectrum $\Delta\rho=\rho_1-\rho_2$, where $\rho_1$ and $\rho_2$ are the two maximum Schmidt eigenvalues of the ladder, $\Delta\rho=0$ means that the entanglement spectrum is doubly degenerate. The SPT phases are characterized by doubly degeneracy in the entanglement spectrum and vanishing of all magnetic orders.}\label{fig:JxJz}
\end{figure}
\begin{figure}[b]
 \centering
 \includegraphics[width=3.2in]{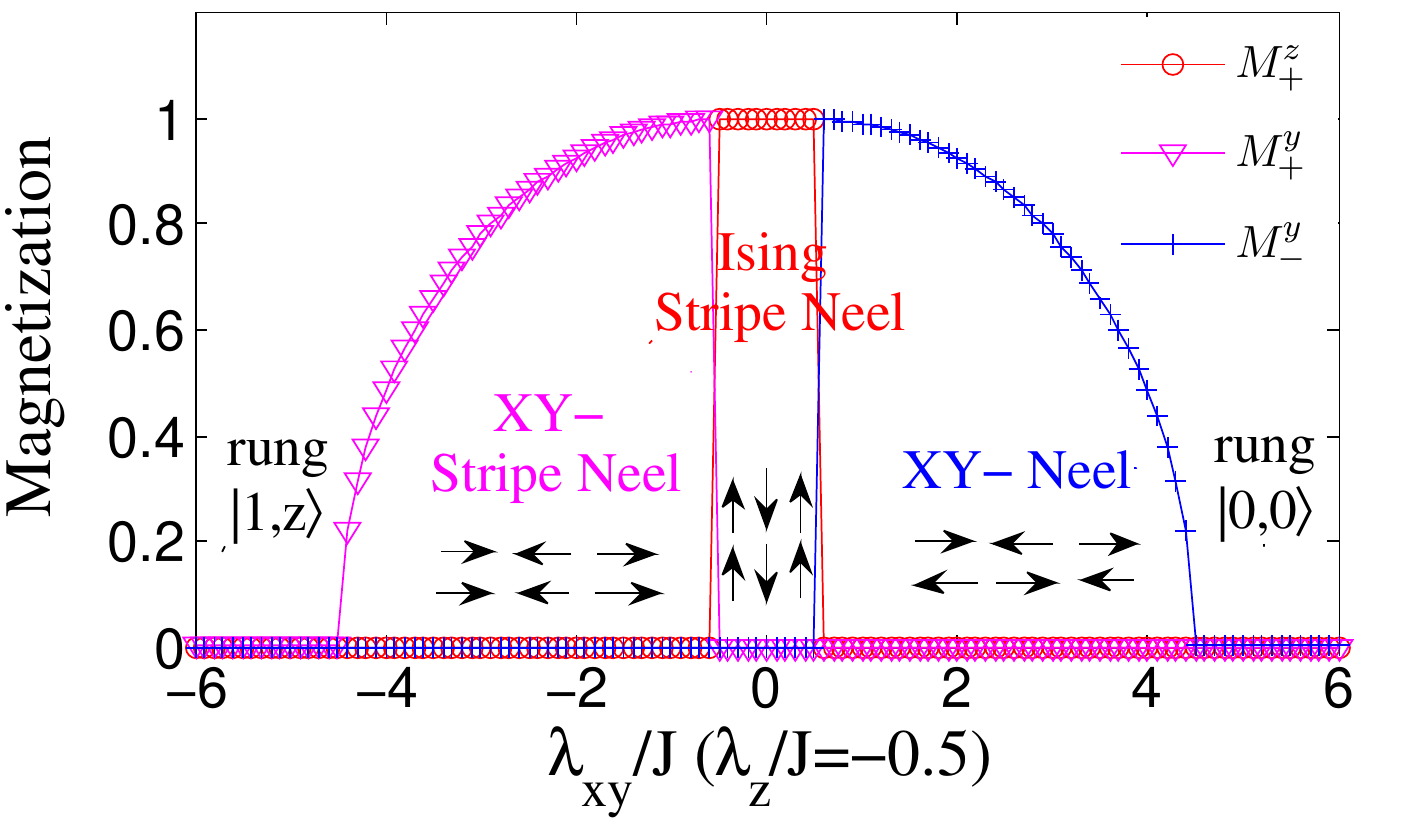}
 \caption{(Color online) The order parameters $M_+^z$ and $M_\pm^{y}$ in the semiclassical approach. Compared to Fig.~\ref{fig:JxJz}, the two SPT phases are missing.}\label{fig:class}
\end{figure}

Comparing the semiclassical phase diagram with Fig.~\ref{fig:JxJz}, we find that the two phase diagrams are
similar except for the absence of two SPT phases in Fig.~\ref{fig:class}.
Importantly, the locations of the quantum SPT phases are close to the phase
boundaries between the Ising ordered phase ($M_+^z\neq0$) and the $xy$-planar ordered phases ($M_\pm^y\neq0$)
of the semiclassical phase diagram. Similar situations occur in spin-1 XXZ
chain, where a Haldane phase exists between
two ordered phases.~\cite{GuWen09,String} This suggests that we can roughly obtain a quantum phase
diagram with a semiclassical approach by `inserting' a SPT phase at the phase
boundary of two different classically ordered phases.

\begin{figure}[t]
 \centering
 \includegraphics[width=3.0in]{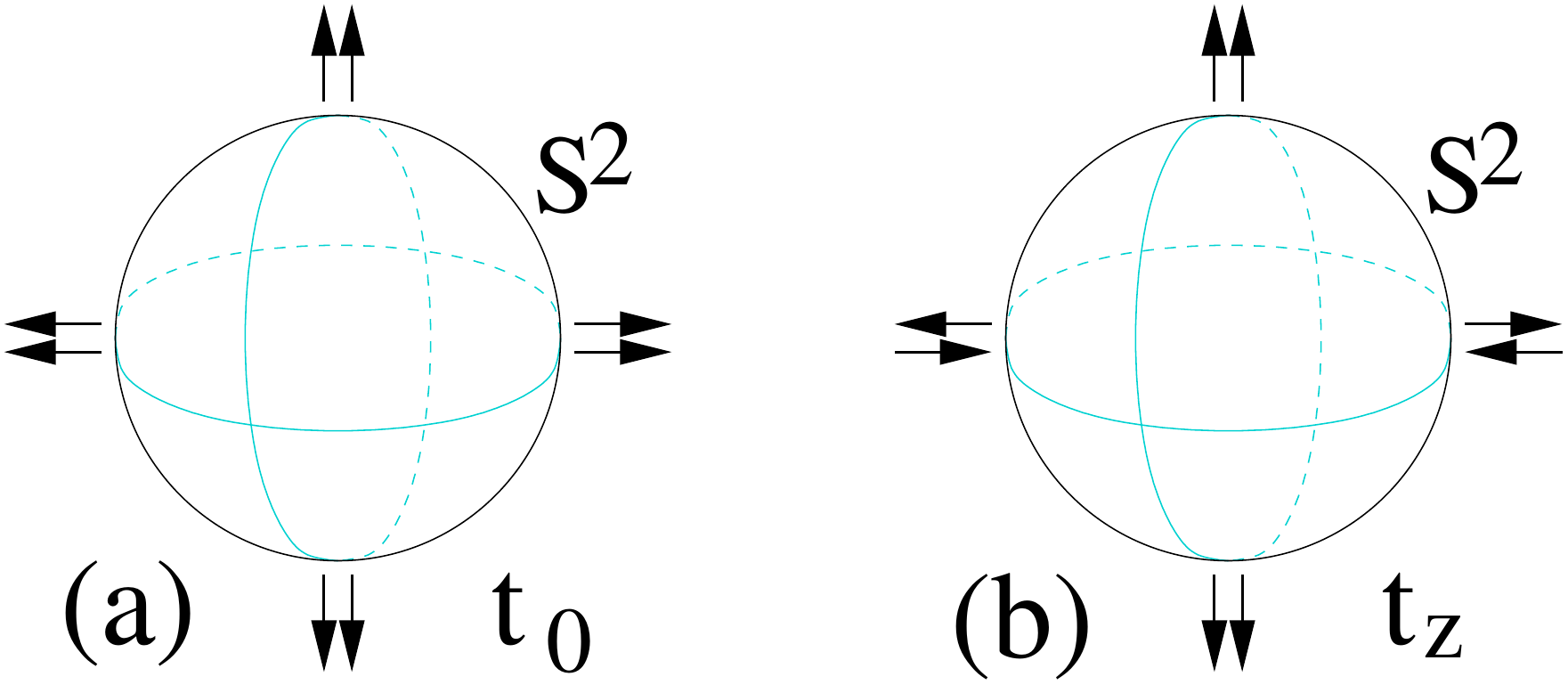}
 \caption{(Color online) (a) At $\lambda_{xy}^c=\lambda_z <0$, the classical
states on the sphere are close in energy. Each direction on the sphere defines a classical state in which two spins at the same rung are parallel and pointing in that direction. (b) At $\lambda_{xy}^c=-\lambda_z>0$, the classical
states on the sphere are also nearly degenerate. But now the two spins at the same rung
are antiparallel in the $x$-$y$ plane and parallel off the plane.
} \label{T0Tz} \end{figure}

Furthermore, the semiclassical picture even indicates some important
information of the quantum SPT phases. For instance, it tells us why the $t_0$
phase is different from the $t_z$ phase. The classical phase boundary
between the two ordered phases is located at $|\lambda_{xy}^c| = |\lambda_{z}|$. At the
point $\lambda_{xy}^c=\lambda_{z}<0$, the states have the same lowest energy if
the spins are antiferromagnetically ordered along the leg and are parallel along the rung polarizing in $y$-$z$ plane (namely, the classical ground states are highly degenerate).
Quantum fluctuations (which is enhanced by the degeneracy) will drive the system into the higher energy states (e.g. some spins are pointing off the $y$-$z$ plane) with a certain weight. As a consequence, the ground state is short-range correlated with a finite excitation gap. Furthermore, edge states exist at each boundary. The edge staes can be considered as an effective spin whose magnetic momentum is half of the total momentum of the two spins at a rung. Since the two spins in the same rung are parallel, the edge states carry free magnetic
moment and respond to magnetic field along arbitrary directions. On the other
hand, at the point $\lambda_{xy}^c= -\lambda_{z}>0$, the lowest energy classical states are those that the two spins are parallel along the $z$ direction and anti-parallel
in the $x$-$y$ plane (see Fig.~\ref{T0Tz}b). Quantum fluctuations around the
sphere drive this state to the $t_z$ phase. In the $t_z$ phase, the effective
edge spins have no net magnetic moment in the $x$-$y$ plane, so they will not
respond to the magnetic field in the $x$-$y$ plane. Similar things happen in the
$t_x$ and $t_y$ phases. Above arguments are applicable to other systems with
different symmetry groups, and provide a guidance to seek SPT phases and
analyze their physical properties.

\subsection{the case $\delta$=1}

\subsubsection{Finite size effect in numerical method}
When establishing the phase diagram Fig.~\ref{fig:phasedg}  and Fig.~\ref{fig:JxJz}, we have approximated the ground state by a matrix product state (MPS), which is obtained through iTEBD method.~\cite{ITEBD} The matrices in the MPS have a finite dimension $D$, which introduces a cutoff to the number of Schmidt eigenvalues in the entanglement Hilbert space. While the energy of the actual ground state can be estimated using a finite-$D$ MPS with high accuracy, the order parameters are usually overestimated due to the finite dimension of the matrix. Furthermore, in some cases the order parameters (which are finite if $D$ is finite) even vanish in the infinite $D$ limit. So a scaling of the order parameters with respect to the dimension $D$ is necessary in order to obtain the correct phase diagram. In the following we will illustrate, via finite size scaling, that the two XY phases disappear if $\delta=1$.

\begin{figure}[b]
  \centering
  \subfigure[]{
    \label{fig:JxJzD=30} 
    \includegraphics[width=3in]{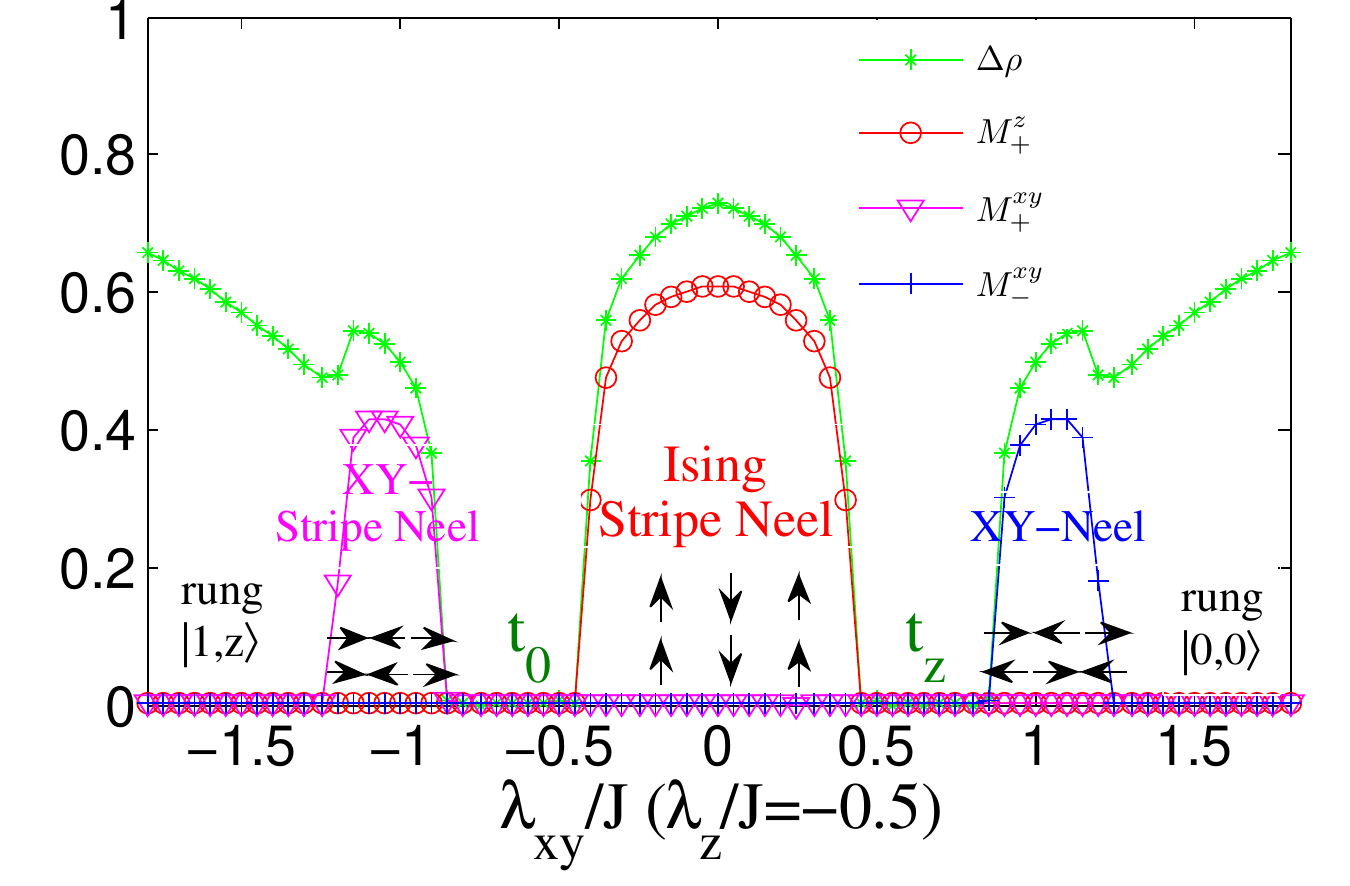}}
  \subfigure[]{
    \label{fig:phasedgrmHeis} 
    \includegraphics[width=2.9in]{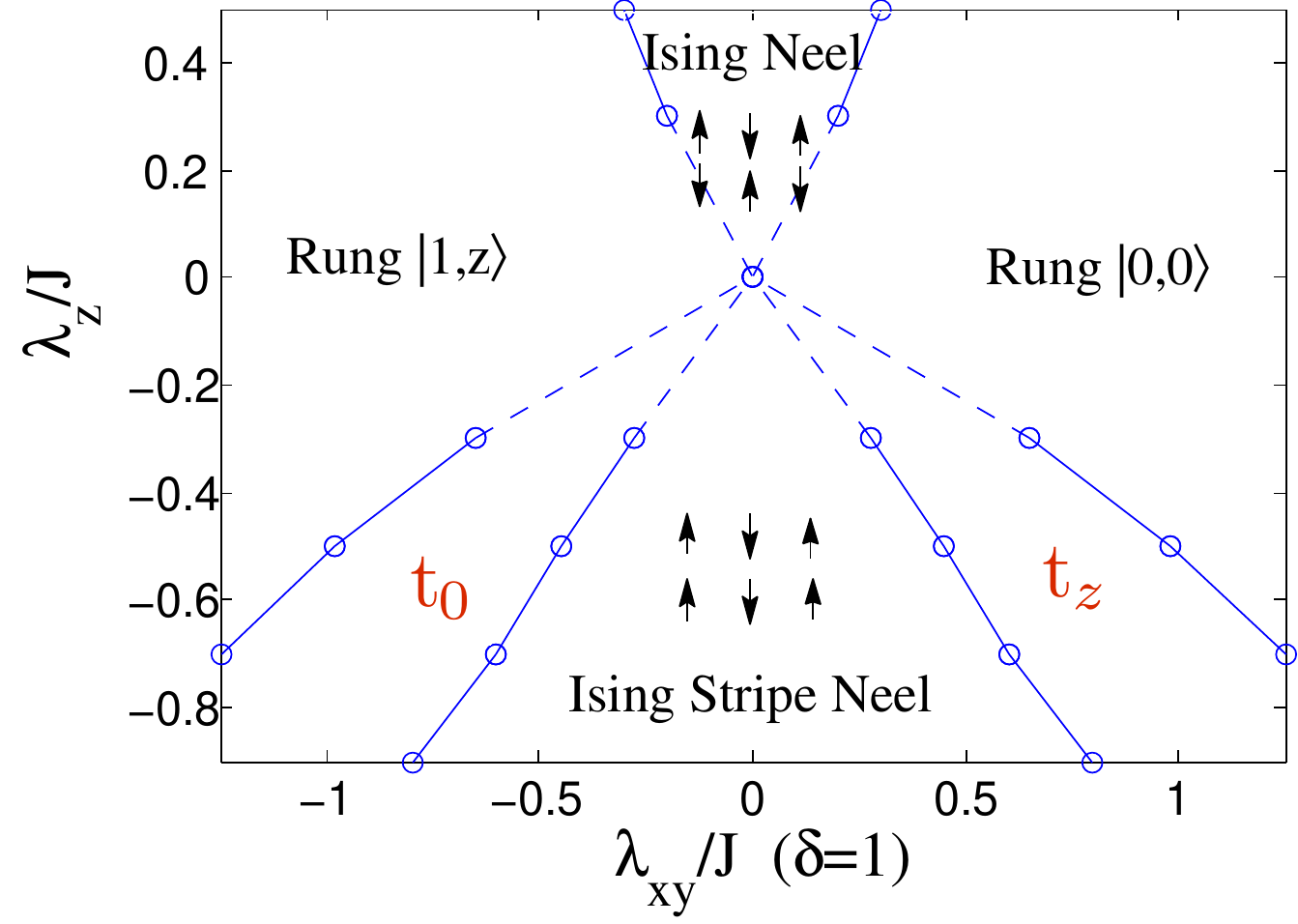}}
  \caption{(Color online) (a)Phase diagram obtained with $D=30$ for $\delta=1,\ \lambda_z=-0.5J$.
The order parameters are defined as $M_+^z=|\langle S_1^z+ S_2^z\rangle|$  and $M_\pm^{xy}=\sqrt{\langle S_1^x\pm S_2^x\rangle^2+\langle S_1^y\pm S_2^y\rangle^2}$;  (b) Phase diagram of $\delta=1$ in the limit $D\to\infty$. }
  \label{fig:phsdgm} 
\end{figure}
\begin{figure}[t]
  \centering
  \subfigure[]{
    \label{fig:Mxyfit} 
    \includegraphics[width=3in]{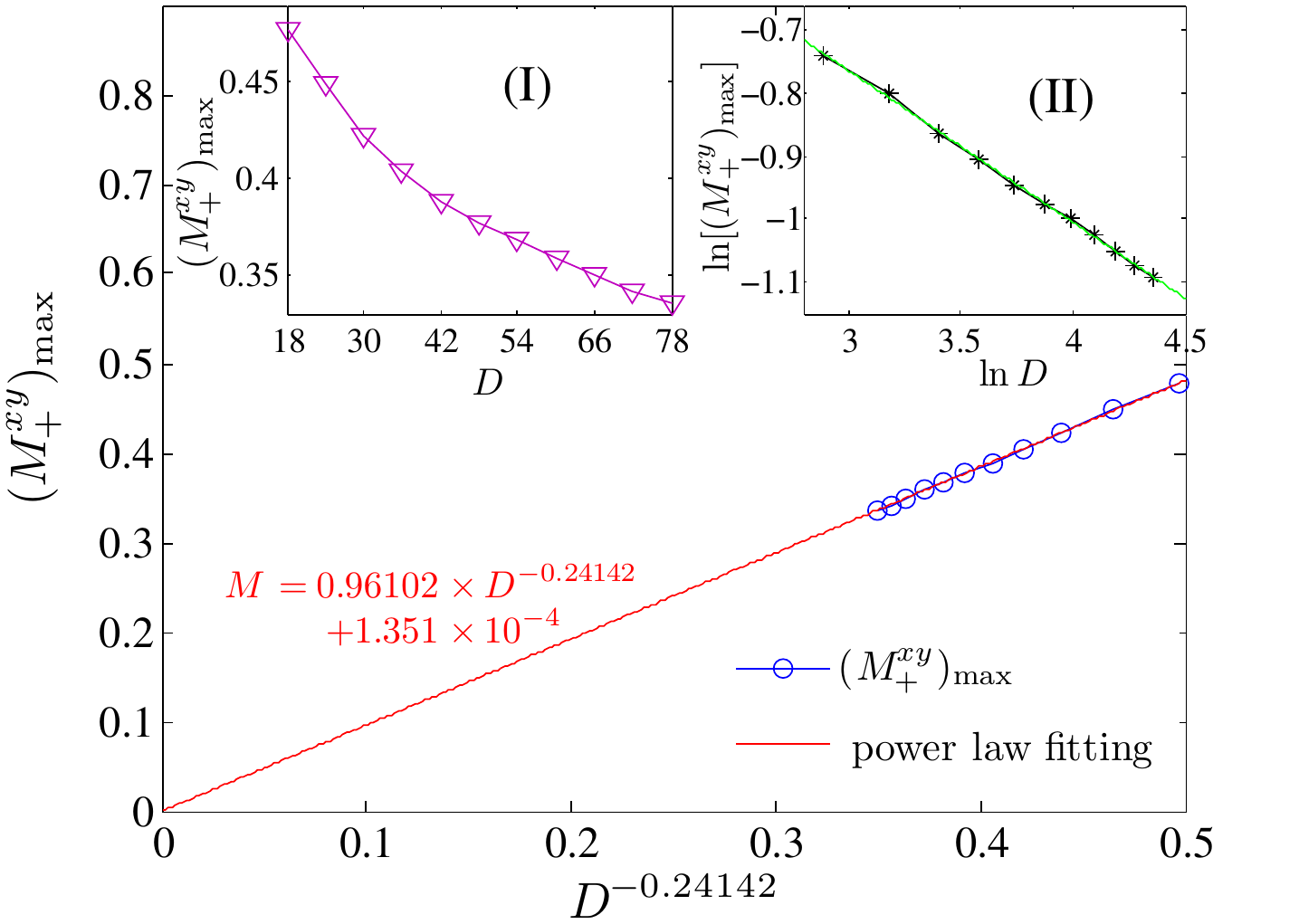}}
  \subfigure[]{
    \label{fig:Mzfit} 
    \includegraphics[width=3in]{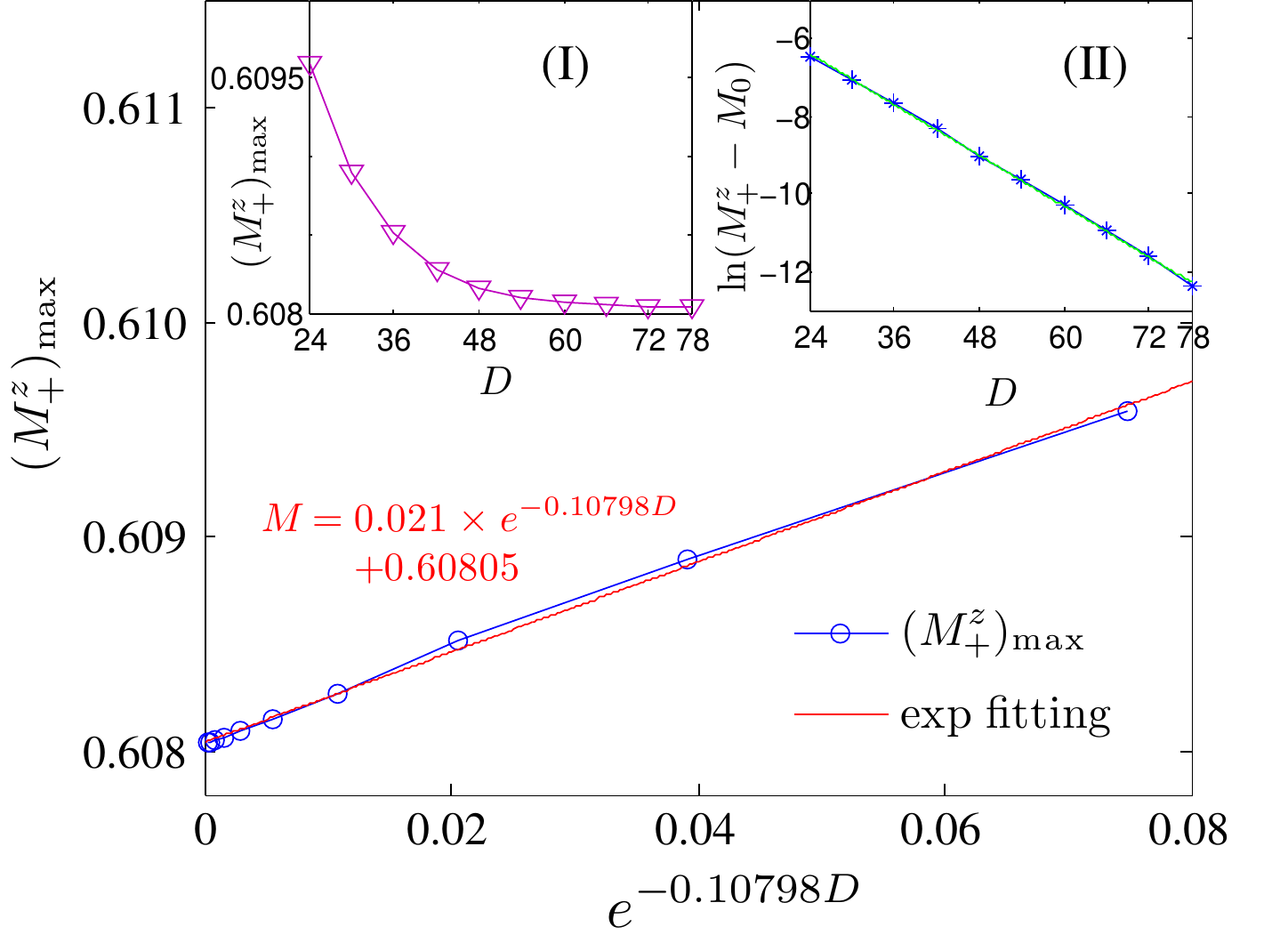}}
  \caption{(Color online) (a)Power law scaling of the maximum value of the order parameter $M_+^{xy}$ with $\delta=1$. The intercept shows that $M_+^{xy}$ is vanishing in power law with the dimension $D$. Insert I shows the data of $M_+^{xy}$ vs. dimension $D$. Insert II is a log-log fit of the data;  (b) Exponential scaling of the maximum value of the order parameter $M_+^{z}$ with $\delta=1$. The intercept shows that $M_+^{z}$ is finite at infinite $D$. Insert I shows the data of $M_+^{z}$ vs. dimension $D$. Insert II is a log fit of the data.}
  \label{fig:HeisFit} 
\end{figure}

The phase diagram in Fig.~\ref{fig:JxJzD=30} with $\delta=1, \lambda_z=-0.5J$ is obtained by setting $D=30$. To evaluate the order parameters of each phase in the infinite $D$ limit, we perform numerical calculations with different $D$'s, as shown in Fig.~\ref{fig:HeisFit}. Performing finite size scaling and extrapolating to infinite $D$, we find that the maximum value of $M_+^{xy}$ in the XY-strip Neel phase (and the same for $M_-^{xy}$ in the XY-Neel phase) vanishes as a power law of $D^{-0.24142}$. In contrast, the maximum value of $M_+^z$ in the Ising stripe Neel phase approaches a finite number exponentially fast. This shows that when $\delta=1$ 
the two XY phases in Fig.~\ref{fig:JxJzD=30} are due to finite size effect (notice that $M_+^{xy}$ decays very slowly with increasing $D$, so any finite $D$ will give incorrect results) and will disappear in the phase diagram in the limit $D\to\infty$ (see Fig.~\ref{fig:phasedgrmHeis}).


%

\begin{figure}[t]
\centering
\includegraphics[width=3.5in]{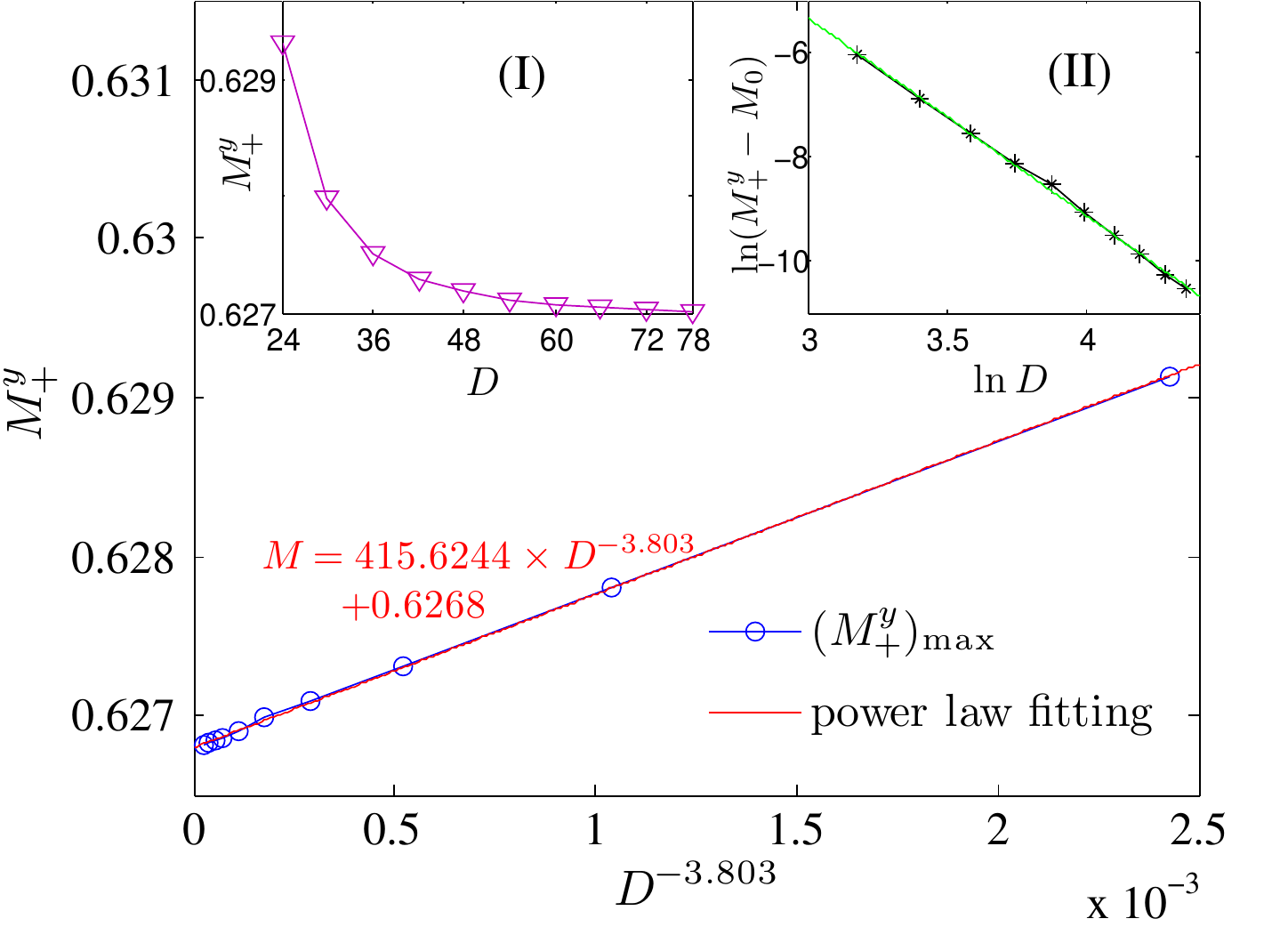}
\caption{(Color online) Power law scaling of the maximum value of the order parameter $M_+^{y}$ with $\delta=0.9$. The intercept shows that $M_+^{y}$ is finite at infinite $D$. Insert (I) shows the data of $M_+^{y}$ VS. dimension $D$. Insert (II) is a log-log fitting of the data.}
\label{fig:XXZMyfit}
\end{figure}

To illustrate that the XY phases will not disappear if $\delta\neq1$, we perform a scaling for the order parameter $M_+^y$ with $D$ for $\delta=0.9$. It turns out that $M_+^y$ still varies as a power law of $D^{-3.803}$, when shifted by a constant $M_0$ (See Fig.~\ref{fig:XXZMyfit}). The fact that $M_0\neq0$ shows that $M_+^y$ in the XY-strip Neel phase (and the same for $M_-^y$ in the XY Neel phase) is finite in the limit $D\to\infty$, and the phase diagram at finite $D$ are qualitatively correct if $\delta\neq1$.

\subsubsection{Vanishing of the XY phases and $U(1)$ symmetry protected phase transitions}
Physically, the disappearance of the two XY phases at $\delta=1$ can be explained by quantum fluctuations.

When $\delta=1$, the Hamiltonian (3) has an enhanced symmetry $[U(1)\rtimes Z_2]\times\sigma$, where the $U(1)$ subgroup means rotation of the spins along $z$-axis, $Z_2$ is generated by a rotation of the spin for $\pi$ around $x$-axis. Since the continuous symmetry $U(1)$ will never spontaneously break in 1D, if the XY phases exist, they will be quasi-long ranged ordered in $x$-$y$ plane and gapless.

However, strong quantum fluctuations gap out these states and drive them into the SPT phases. From the semiclassical approach, the enhanced symmetry results in a larger degeneracy of the semiclassical ground states near $\lambda_{xy}\sim\lambda_z$. The enlarged degeneracy of the semiclassical ground states enhances the quantum fluctuations and gap out all the states (except the states at the critical points). Consequently the XY phases disappear, and the nontrivial SPT phase $t_0/t_z$ touches the trivial phase rung-$|1,z\rangle$/rung-$| 0,0\rangle$ directly. That is to say, the direct transition between $t_0/t_z$ and  rung-$|1,z\rangle$/rung-$| 0,0\rangle$ is protected by the continuous $U(1)$ symmetry.

Notice that similar situations also happen in $S=1$ chains,~\cite{GuWen09} where the Haldane phase and the trivial phase (in analogy to the rung-phases) are separated by a $Z_2$ symmetry breaking phase when the system does not have a continuous symmetry [such as $U(1)$ spin rotational symmetry].  

\section{Physical realization and quantum simulation}

Here we propose possible realizations of the SPT phases in two-legged ladder Mott systems.~\cite{Mott} The Hamiltonian
(\ref{lambda}) can be considered as two Heisenberg chains coupled with magnetic
dipole-dipole interaction and exchange interaction $H_T=\sum_i[\eta(\mathbf
S_{1,i}\cdot\mathbf S_{2,i})+\gamma(\mathbf S_{1,i}\cdot\mathbf S_{2,i}- 3
S_{1,i}^zS_{2,i}^z)]$ with $\lambda_z=\eta-2\gamma$ and
$\lambda_{xy}=\eta+\gamma$. However, in real materials, the magnetic
dipole-dipole interaction is too weak to support the SPT phases. On the other
hand, axial anisotropy interaction $D_z(S_1^z+S_2^z)^2= 2D_z(S_1^zS_2^z+1)$ can
yield the Hamiltonian (\ref{lambda}) by $H_T=\sum_i[\eta(\mathbf
S_{1,i}\cdot\mathbf S_{2,i})+D_z(S_{1,i}^z+S_{2,i}^z)^2]$ with
$\lambda_z=2D_z+\eta$, $\lambda_{xy}=\eta$. Note that a weak effective spin-1
axial anisotropy term exists in the material $(\mathrm C_5\mathrm H_{12}\mathrm
N)_2\mathrm {CuBr}_4$.~\cite{anisotropy} If this term is
negative and is strong enough in some material, then the $t_z$ phase will be
realized.

\begin{figure}[t]
\includegraphics[width=3.3in]{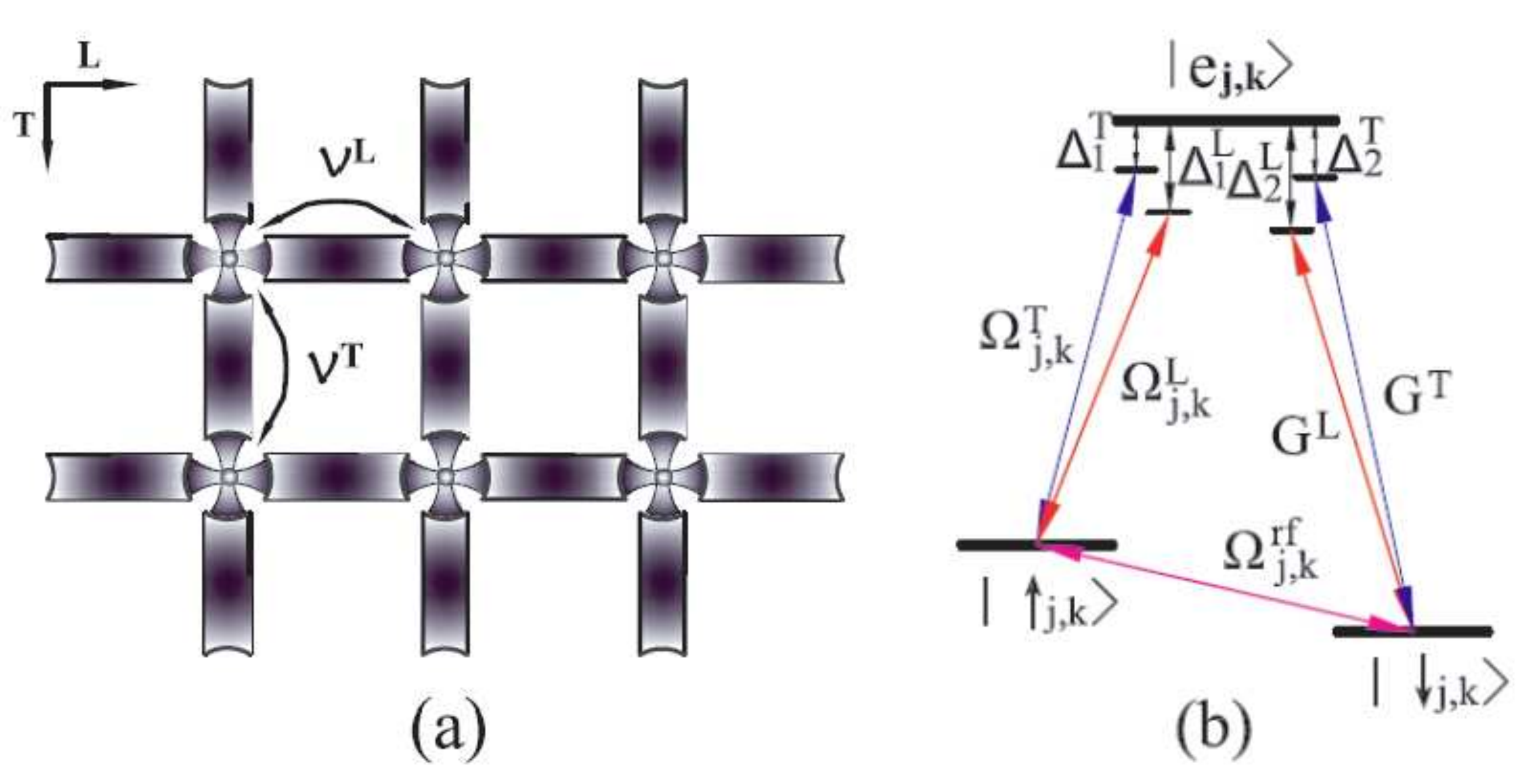}
\caption{(Color online) (a) Illustration of the coupled-cavity ladder. A single atom is held at the intersection of two cavity modes that are coupled across both the longitudinal ($L$) and the transverse ($T$) directions, with coupling rates $v^L$ and $v^T$, respectively. (b) Schematic of the coupling scheme for the atom inside the cavity. The external lasers are applied perpendicularly to the plane of the ladder. The subscripts $j,k$ of the parameters indicate the atom is on the $j$th site transversally and $k$th site longitudinally.}\label{figimplement}
\end{figure}

On the other hand, due to the nearest-neighbor-only interactions, it is tempting to consider the possibility of simulating these Hamiltonians in a non-condensed-matter setting. Here we provide a proof-of-principle implementation scheme for the Hamiltonians for the $t_0$ phase (\ref{t0H}) and for the $t_z$ phase (\ref{Hz}) based on coupled-harmonic-oscillator array. Here we only consider the isotropic case with $\delta =1$. Note that it is also possible to implement the more general anisotropic cases with additional Raman lasers along the ladder. Our scheme may be extended to systems including solid-spins interacting with arrays of coupled transmission line resonators, or ions in a Coulomb crystal.  As a concrete example, we illustrate the scheme using a coupled-QED-cavity ladder (see Fig. \ref{figimplement}(a)), where the quantized cavity fields couple to their nearest neighbors in the longitudinal (L) and the transverse (T) directions via photon hopping.~\cite{HartmannPlenio} The spin degrees of freedom on each site are encoded in the internal states of a single atom trapped inside the cavity.

In Fig. \ref{figimplement}(b), we show the coupling scheme for a minimal model with three internal states, with $\left(|\uparrow_{j,k}\rangle,|\downarrow_{j,k}\rangle\right)$ corresponding to hyperfine states in the ground state manifold and $|e_{j,k}\rangle$ corresponding to an electronically excited state. Two far-detuned external laser fields with Rabi frequencies $\Omega _{j,k}^\mu =\Omega ^\mu e^{i\theta _{j,k}^\mu }$ ($\mu =L,T$) couple $\left| \uparrow _{j,k}\right\rangle $ and $\left| e_{j,k}\right\rangle $, while two far-detuned cavity modes with Rabi frequencies $G^\mu $ couple $\left| e_{j,k}\right\rangle $ and $\left| \downarrow _{j,k}\right\rangle $, with all the detunings satisfying $\left|\Delta _2^L-\Delta _2^T\right|,$ $\left|\Delta _1^L-\Delta _1^T\right|,\left|\Delta _1^\mu\right| ,\left|\Delta _2^\mu\right| \gg G^\mu ,\Omega ^\mu ,\left|\Delta _2^\mu -\Delta _1^\mu\right| $. Hence, we have two independent Raman paths in the longitudinal and in the transverse directions, and a virtually populated excited state $|e_{j,k}\rangle$. Finally, the hyperfine states are coupled using a resonant radio-frequency (r.f.) dressing field with Rabi frequency $\Omega^{\mathrm{rf}}_{j,k}=\Omega^{\mathrm{rf}}e^{i\varphi_{j,k}}$. One can show that all the cavity modes are virtually excited under the condition $\left| \Delta _2^\mu -\Delta _1^\mu \right| \sim  \Omega ^{\mathrm{rf}} \gg g^\mu/\sqrt{2N},v^\mu$, where $g^{\mu}=G^{\mu}\Omega ^\mu (\Delta_1^{\mu}+\Delta_2^{\mu})/2\Delta_1^{\mu}\Delta_2^{\mu}$, $v^{\mu}$ is the coupling rate between cavities along the direction $\mu$, and $N$ is the number of sites in the longitudinal direction.~\cite{ZhengGuo} After adiabatic elimination of the excited states and the cavity modes, the resulting effective Hamiltonian gives rise to effective spin-spin interactions between the neighboring sites, with the pseudo-spins in the effective Hamiltonian given by the
superpositions of the hyperfine states: $|\uparrow_{j,k}'\rangle=\left(e^{i\varphi_{j,k}}|\uparrow_{j,k}\rangle+|\downarrow_{j,k}\rangle\right)/\sqrt{2}$, $|\downarrow_{j,k}'\rangle=\left(-|\uparrow_{j,k}\rangle+e^{-i\varphi_{j,k}}|\downarrow_{j,k}\rangle\right)/\sqrt{2}$.

Depending on the magnitude and the relative phases of the coupling fields, the effective Hamiltonian can support either the $t_0$ phase or the $t_z$ phase. In particular, with $\Omega^{\mathrm{rf}}\sim 0.75{\left|\Delta^{\mu}_2-\Delta^{\mu}_1\right|}$ and $\left| \theta ^{L}_{j,k+1}-\theta^{L} _{j,k}\right| =\pi $, we get the Hamiltonian for the $t_0$ phase when $\theta_{j,k}^T=\varphi_{j,k}=0$ for arbitrary $j,k$; and we get the Hamiltonian for the $t_z$ phase when $\left|\theta^T_{1,k}-\theta^T_{2,k}\right|=\left| \varphi _{1,k}-\varphi _{2,k}\right| =\pi $ and $\varphi_{j,k}=\varphi_{j,k+1}$ for arbitrary $j,k$. These conditions can be achieved by modulating the relative phases of the external coupling fields so that they are either $0$ or $\pi$ between neighboring sites. The interaction rates $J$ and $\lambda$ in Hamiltonian (\ref{t0H},\ref{Hz}) are given as $J \sim 1.14\left(g^L\right)^2v^L/\left(\Omega^{\mathrm{rf}}\right)^2$, $\lambda = -1.14\left(g^T\right)^2v^T/\left(\Omega^{\mathrm{rf}}\right)^2$.  For typical experimental parameters \cite{BoozerKimble}: $\Omega ^\mu\sim 100$MHz, $G^{\mu}\sim 100$MHz, $\left|\Delta^{\mu}_i\right|\sim 1$GHz ($i=1,2$), $v^{\mu}\sim10$MHz, $\Omega^{\mathrm{rf}}\sim 100$MHz, we have $J\sim 0.11$MHz, with the magnitude of $\lambda/J$ widely tunable by adjusting the ratio between $\Omega^L$ and $\Omega^T$.

The single-site addressability of the atoms in the cavity ladder allows much freedom in probing the properties of the system.  For the SPT phases in which we are interested, we need to measure the response of the edge states to the magnetic field. As the pseudo-spins are related to the hyperfine states via a rotation, we may implement an effective magnetic field on the pseudo-spins by imposing the appropriately rotated operations on the edge states. For instance, to apply an effective magnetic field in the $z$-direction, we need to apply a $\sigma_x$ rotation on the hyperfine states (assuming $\varphi_{j,k}=0$), which can be achieved for example by adding a resonant r.f. field between the hyperfine states, with the effective Rabi frequency corresponding to the magnitude of the effective magnetic field. To probe the response of the edge states to the effective field, we may measure the polarization of the ground state of the system. Alternatively, as the effective magnetic field may induce an energy splitting between the spin states, we can measure the existence of energy splitting as a response of the SPT phases to the external field. As an example, for the $t_z$ phase under an effective magnetic field along $z$, we need to implement the following steps on the two edge sites at one end of the open boundaries: (i) adiabatically turn on resonant r.f. fields between the hyperfine states so that the degeneracy of the edge states is lifted and an energy splitting appears between $|\uparrow'_{\text{edge}}\rangle$ and $|\downarrow'_{\text{edge}}\rangle$; (ii) in the presence of the effective magnetic field, apply an effective resonant coupling fields between the pseudo-spins, which corresponds to two-photon detuned Raman fields between the hyperfine states with the Stark shifts equal to the Rabi-frequency of the effective resonant coupling fields between the pseudo-spins; (iii) after some time of evolution, rapidly turn off all the coupling fields, then apply a $\pi/2$-pulse on the hyperfine states, so that pseudo-spin population is projected onto that of the hyperfine states; (iv) the population of the hyperfine state can be probed for example by high-fidelity hyperfine state readout technique based on cavity-enhanced fluorescence.~\cite{detection} If the edge states are responsive to the effective magnetic field, one will observe Rabi-oscillations in the measured fluorescence. Measuring the magnetic field response in all three spatial directions will allow us to establish the signature of the $t_z$ phase.
\\

\section{Conclusion and Acknowledgement}

In summary, we have studied nontrivial quantum phases $t_0$(the Haldane phase) and $t_z$(a new SPT phase different from the Haldane phase) protected by $D_2\times\sigma$ symmetry in a spin-1/2 ladder model. The model has simple two-body interactions and a rich phase diagram. We then provided a semiclassical understanding of the physical properties of the SPT phases, and discussed the general principles for the search of these novel phases. Finally, we have proposed a proof-of-principle quantum simulation scheme of nontrivial SPT phases in cold atom systems.

ZXL would like to thank Fa Wang, Xie Chen, Tian-Heng Han and Salvatore R. Manmana for helpful discussions. This work is supported by NFRP (2011CB921200, 2011CBA00200), NNSF (60921091), NSFC (11105134,11105135), the Fundamental Research Funds for the Central Universities (WK2470000001, WK2470000004, WK2470000006). XGW is supported by NSF DMR-1005541 and NSFC 11074140.

\appendix
\section{Constructing the Hamiltonians with SPT phases}\label{App: Hamiltonain}
In this Appendix, we demonstrate the method~\cite{LCW} by which we obtain the Hamiltonian $H_0$, $H_x$, $H_y$ and $H_z$ in the main text. The procedure contains three steps: (i) construct a matrix product state (MPS) wave function with given edge states which are described by the projective representations; (ii) construct the parent Hamiltonian for the MPS using projection operators; (iii) simplify the parent Hamiltonian by adiabatic deformations. In the first step, we need to know the projective and linear representations of the symmetry group, together with the Clebsch-Gordan (CG) coefficients. In the following, we will first present necessary information and then discuss the construction of the Hamiltonian step-by-step.

The four projective representations of $D_2\times\sigma$ corresponding to the $t_0$,$t_x$,$t_y$,$t_z$ phases are given in Tab.~\ref{tab:SPT_D2P} (we only provide the representation matrices for the generators), and the eight linear representations are listed in Tab.~\ref{tab:linear}.
\begin{table}[htbp]
\caption{Four nontrivial SPT phases in $S=1/2$ spin ladders respecting $D_2\times\sigma$ symmetry.}
\label{tab:SPT_D2P}
\begin{ruledtabular}
\begin{tabular}{c|ccc|c}
&$M(R_z)$&$M(R_x)$&$M(\sigma)$&active operators\\
\hline
$E_3$  ($t_0)$       & $i\sigma_z$& $ \sigma_x$&     I      &$(S^x_{+},S^y_+,S^z_+)$ \\
$E_5$  ($t_x)$       & $i\sigma_z$& $ \sigma_x$& $i\sigma_x$&$(S^x_{+},S^y_-,S^z_-)$ \\
$E_6$  ($t_z)$       & $i\sigma_z$& $i\sigma_x$& $ \sigma_z$&$(S^x_{-},S^y_-,S^z_+)$\\
$E_7$  ($t_y)$       & $i\sigma_z$& $i\sigma_x$& $i\sigma_y$&$(S^x_{-},S^y_+,S^z_-)$ 
\end{tabular}
\end{ruledtabular}
\end{table}

\begin{widetext}
\begin{center}
\begin{table}[htbp]
\caption{Linear representations of $D_2\times \sigma$. The four bases are defined as $|0,0\rangle={1\over\sqrt2}(|\uparrow_1
\downarrow_2\rangle-|\downarrow_1 \uparrow_2\rangle)$, $|1,x\rangle ={1\over\sqrt2}(|\downarrow_1 \downarrow_2\rangle-|\uparrow_1
\uparrow_2\rangle)$, $|1,y\rangle={i\over\sqrt2} (|\downarrow_1\downarrow_2 \rangle +|\uparrow_1 \uparrow_2\rangle)$, and $|1,z\rangle ={1\over\sqrt2} (|\uparrow_1 \downarrow_2\rangle+ |\downarrow_1 \uparrow_2\rangle)$, where the subscripts $1,2$ label the different spins on the same rung. The operator $S^m_+= S^m_1+S^m_2$ ($m=x,y,z$) is even under inter-chain reflection, and $S^m_-=S^m_1 -S^m_2$ is odd under the reflection.}
\label{tab:linear}
\begin{ruledtabular}
\begin{tabular}{c|ccc|c|cc}
            &  $R_z$    &$R_x$        & $\sigma$   &bases&operators&\\
\hline
$A_{g}$     &      1    &      1      &  1    & & & \\
$B_{1g}$    &      1    &     -1      &  1    &$|1,z\rangle$&$S^{z}_{+}$& \\
$B_{2g}$    &     -1    &     -1      &  1    &$|1,y\rangle$&$S^{y}_{+}$& \\
$B_{3g}$    &     -1    &      1      &  1    &$|1,x\rangle$&$S^{x}_{+}$& \\
\hline
$A_{u}$     &      1    &      1      & -1    & $|0,0\rangle$  &&\\
$B_{1u}$    &      1    &     -1      & -1    & &$S^{z}_{-}$&\\
$B_{2u}$    &     -1    &     -1      & -1    & &$S^{y}_{-}$& \\
$B_{3u}$    &     -1    &      1      & -1    & &$S^{x}_{-}$& 
\end{tabular}
\end{ruledtabular}
\end{table}
\end{center}
The Hilbert space of the direct product of two projective representations can be reduced to a direct sum of linear representations. The CG coefficients (we assume that all the CG coefficients are real numbers) are :
\begin{eqnarray}\label{CG}
&&E_3\otimes E_3=A_g\oplus B_{1g}\oplus B_{2g}\oplus B_{3g},\ \ C^{A_g}=\sigma_x,\ C^{B_{1g}}=i\sigma_y,\ C^{B_{2g}}=\sigma_z,\ C^{B_{3g}}=I;\\
&&E_5\otimes E_5=B_{1g}\oplus B_{2g}\oplus A_u\oplus B_{3u},\ \ C^{A_u}=\sigma_x,\ C^{B_{3u}}=I,\ C^{B_{1g}}=i\sigma_y,\ C^{B_{2g}}=\sigma_z;\\
&&E_{6}\otimes E_{6}=B_{2g}\oplus B_{3g}\oplus A_u\oplus B_{1u},\ \ C^{A_u}=i\sigma_y,\ C^{B_{1u}}=\sigma_x,\ C^{B_{2g}}=I,\ C^{B_{3g}}=\sigma_z;\\
&&E_{7}\otimes E_{7}=A_{g}\oplus B_{2g}\oplus B_{1u}\oplus B_{3u},\ \ C^{A_{g}}=i\sigma_y,\ C^{B_{2g}}=I,\ C^{B_{1u}}=\sigma_x,\ C^{B_{3u}}=\sigma_z,
\end{eqnarray}
where  $|m\rangle=C^m_{\alpha\beta}|\alpha\rangle|\beta\rangle$, $|m\rangle$ is the basis of a linear representation and $|\alpha\rangle,|\beta\rangle$ are the bases of two projective representations.

In the following,  we will illustrate the method to obtain the Hamiltonian of the $t_0$ phase as an example.

The first step is obtaining the MPS. From Tab.~\ref{tab:SPT_D2P}, the edge states of the $t_0$ phase are described by the $E_3$ projective representation. In an ideal MPS, every rung is represented by a direct product of two $E_3$ projective representations, which can be reduced to four linear representations, $E_3\otimes E_3=A_g\oplus B_{1g}\oplus B_{2g}\oplus B_{3g}$. From Tab.~\ref{tab:linear}, $B_{1g}, B_{2g}, B_{3g}$ correspond to the bases $|1,z\rangle, |1,y\rangle, |1,x\rangle$ respectively. The basis $A_g$ (or $|0,0,\rangle$) is absent on every rung in the MPS state. Thus, the support space for the ideal MPS is the Hilbert subspace $\otimes_i(|1,z\rangle \oplus|1,y\rangle \oplus|1,x\rangle)_i$, where $i$ is the index of rung. From the CG coefficients (\ref{CG}), we can write such an ideal MPS which is invariant (up to a phase) under the symmetry group:
\[|\psi\rangle=\sum_{\{m_1,...,m_N\}}\mathrm{Tr}(A^{m_1}...A^{m_N})|m_1...m_N\rangle,\]
with $A^m=e^{i\theta_m}BC^{m}$. Here $B$ is the CG coefficients of decomposing the product representations $E_3\otimes E_3$ into a 1D representation (here we choose $B=C^{A_g}$), and $e^{i\theta_m}$ can be absorbed into the spin bases. Now we have,
\begin{eqnarray}
A^{|1,x\rangle}=\sigma_x,\ \ A^{|1,y\rangle}=\sigma_y, \ \
A^{|1,z\rangle}=\sigma_z,
\end{eqnarray}

The second step is constructing the parent Hamiltonian, which is a sum of projectors. Each projector is a projection onto the ground state subspace of two neighboring rungs. Assuming the orthonormal bases for the MPS state of two neighboring rungs $i,i+1$ are $\psi_1$,$\psi_2$,$\psi_3$,$\psi_4$, then the projector is $P_{i,i+1}=-(\sum_{a=1}^4|\psi_a\rangle\langle\psi_a|)_{i,i+1}$ and the resultant parent Hamiltonian $H_{0\mathrm{ex}}=\sum_iP_{i,i+1}$ is given as,
\begin{eqnarray}\label{EX}
H_{0\mathrm{ex}}&=&J\sum_i\left[{5\over12}(\mathbf S_{1,i}+\mathbf S_{2,i})\cdot(\mathbf S_{1,i+1}+\mathbf S_{2,i+1})
-{2\over3}\mathbf S_{1,i}\cdot\mathbf S_{2,i}-{2\over3}(\mathbf S_{1,i}\cdot\mathbf S_{2,i})(\mathbf S_{1,i+1}\cdot\mathbf S_{2,i+1})\right.\nonumber\\&&\left.+{1\over3}(\mathbf S_{1,i}\cdot\mathbf S_{1,i+1})(\mathbf S_{2,i}\cdot\mathbf S_{2,i+1})+{1\over3}(\mathbf S_{1,i}\cdot\mathbf S_{2,i+1})(\mathbf S_{2,i}\cdot\mathbf S_{1,i+1})\right].
\end{eqnarray}

The final step is deforming the Hamiltonian. It can be shown that only the first two terms in (\ref{EX}) are important. To see this, we introduce the parameter $d$,
\begin{eqnarray}\label{H}
H&=&J\sum_i\left[{5\over12}(\mathbf S_{1,i}\cdot\mathbf S_{1,i+1}+\mathbf S_{2,i}\cdot\mathbf S_{1,i+1})-{2\over3}\mathbf S_{1,i}\cdot\mathbf S_{2,i}\right]+d\sum_i\left[{5\over12}(\mathbf S_{1,i}\cdot\mathbf S_{2,i+1}+\mathbf S_{2,i}\cdot\mathbf S_{1,i+1})\right.\nonumber\\&&\left.-{2\over3}(\mathbf S_{1,i}\cdot\mathbf S_{2,i})(\mathbf S_{1,i+1}\cdot\mathbf S_{2,i+1})+{1\over3}(\mathbf S_{1,i}\cdot\mathbf S_{1,i+1})(\mathbf S_{2,i}\cdot\mathbf S_{2,i+1})+{1\over3}(\mathbf S_{1,i}\cdot\mathbf S_{2,i+1})(\mathbf S_{2,i}\cdot\mathbf S_{1,i+1})\right].
\end{eqnarray}

Note that when $d/J=1$, (\ref{H}) is the same as (\ref{EX}). Now we study the ground state energy and entanglement spectrum (through time-evolving block decimation method) to see if there is a phase transition when $d$ is varied.
\begin{figure}[htbp]
  \centering
  \subfigure[]{
    \label{fig: link_energy} 
    \includegraphics[width=2.64in]{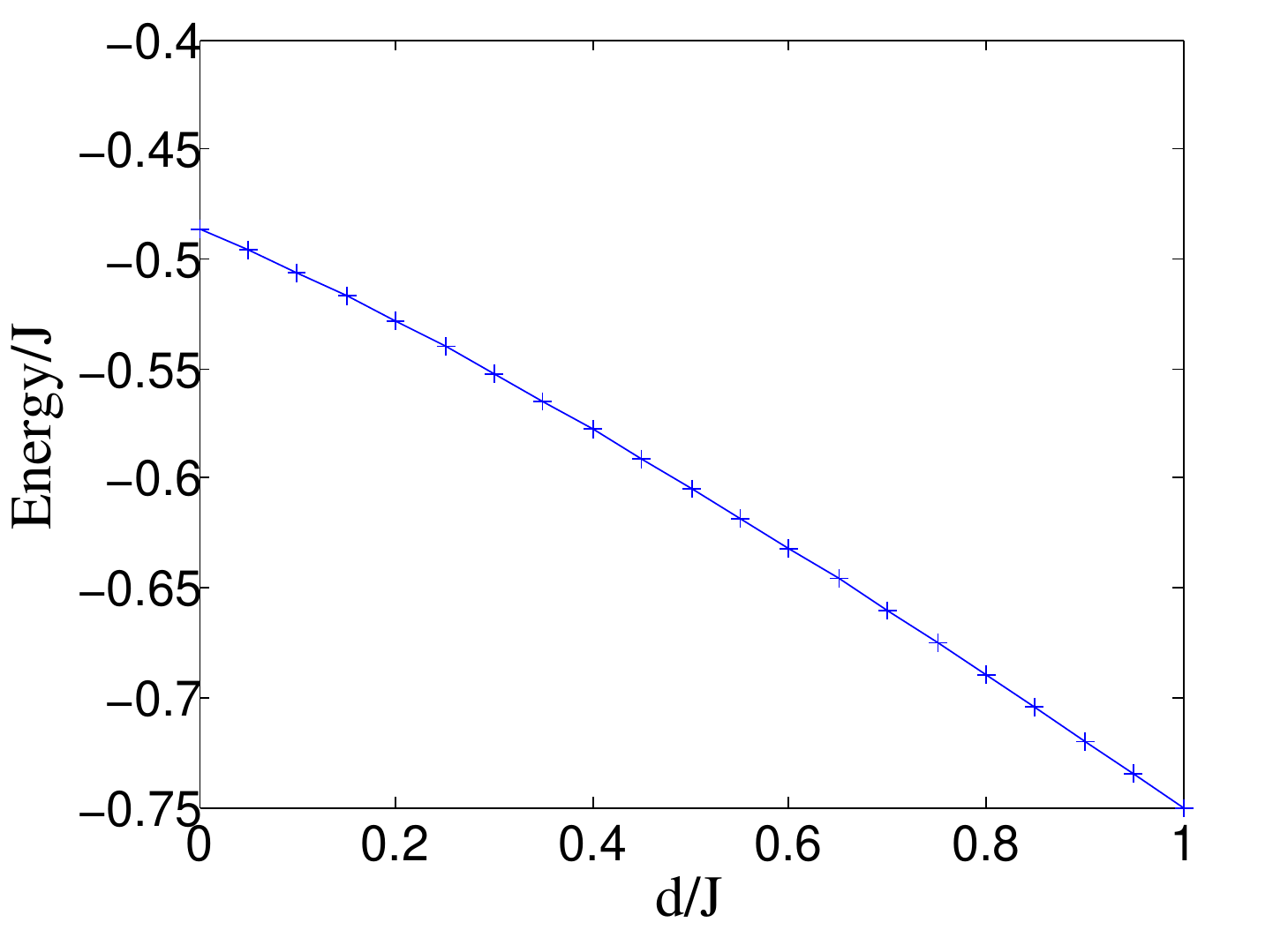}}
  \subfigure[]{
    \label{fig:class_magnetization} 
    \includegraphics[width=2.64in]{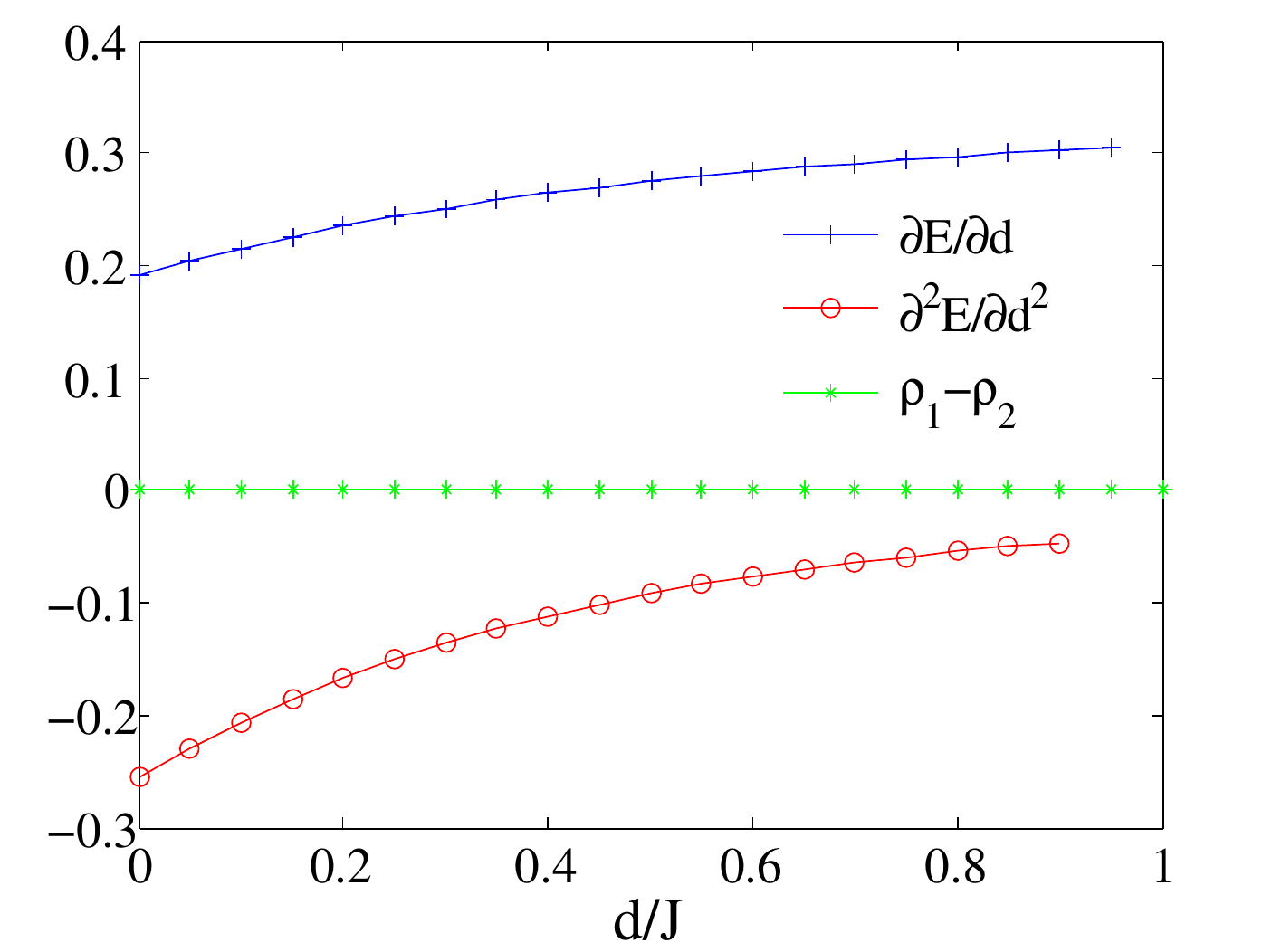}}
  \caption{(Color online) (a) Energy VS. $d/J$.  (b) The derivatives of the energy curve with respect to the parameter $d$ and the information of entanglement spectrum.}
  \label{fig:Link} 
\end{figure}

Fig.~\ref{fig:Link} shows that the energy is a smooth function of parameter $d/J\in[0,1]$. Furthermore, the entanglement spectrum remains degenerate in $d/J\in[0,1]$. Thus we only need nearest-neighbor exchanges to realize the Haldane ($t_0$) phase, which leads to the Hamiltonian $H_0$ in the main text.

The active operators in Tab.~\ref{tab:SPT_D2P} are obtained as the following. In the Hilbert space spanned by the two-fold degenerate edge states, only the Pauli matrices $(\sigma_x,\sigma_y,\sigma_z)$ can lift the degeneracy. But these operators are not physical quantities. We need to find physical spin operators which vary in the same way as these Pauli matrices under the symmetry group. In other words, we require that the active operators form the same linear representations as $(\sigma_x,\sigma_y,\sigma_z)$, respectively. For example, in the $E_3$ projective representation,  $\sigma_x$ varies as
\begin{eqnarray*}
&&M(R_z)^\dag \sigma_x M(R_z)=-\sigma_x,\\
&&M(R_x)^\dag \sigma_x M(R_x)= \sigma_x,\\
&&M(\sigma)^\dag \sigma_x M(\sigma)= \sigma_x.
\end{eqnarray*}
On the other hand, from Tab.~\ref{tab:linear},
\begin{eqnarray*}
&&R_z ^\dag S_+^x  R_z =-S_+^x,\\
&&R_x ^\dag S_+^x  R_x = S_+^x,\\
&&\sigma ^\dag S_+^x  \sigma = S_+^x.
\end{eqnarray*}
We find that $S_+^x$ and $\sigma_x$ belong to the same linear representation $B_{3g}$ under the symmetry operation. This means that in the low energy limit (i.e. in the ground state subspace), these two operators have similar behavior. So we can identify $S_+^x$ as an active operator. Similarly, he operators $S_+^y$ and $S_+^z$ are active operators corresponding to $\sigma_y$ and $\sigma_z$ respectively.

Similar to (\ref{EX}), we can construct the exactly solvable Hamiltonian $H_{z\rm{ex}}$ of the $t_z$ phase. It is the same as (\ref{EX}) except that every $\mathbf A\cdot\mathbf B$ term is replaced by $A_zB_z-A_xB_x-A_yB_y$. This Hamiltonian can be simplified into the form of $H_z$ [Eqn.(\ref{lambda}) of the main text] without any phase transition ($H_x$ and $H_y$ are obtained similarly). 

The active operators in $t_z$ phase can be easily obtained: $S_-^x, S_-^y, S_+^z$. Notice that $S_+^x, S_+^y$ are not active operators, meaning that the edge states in the ground state will not respond to the uniform magnetic in $x$ and $y$ directions. To check this result, we perform a finite-size exact diagonalization of the solvable model $H_{z\rm{ex}}$. As shown in Fig.\ref{fig:BinTz}, only the magnetic field along $z$ direction can split the ground state degeneracy. These properties are valid in the whole $t_z$ phase in  thermodynamic limit. This verifies the conclusion that only $S_+^z$ is the active operator.

\begin{figure}[htbp]
  \centering
  \subfigure[]{
    \label{fig: link_energy} 
    \includegraphics[width=2.25in]{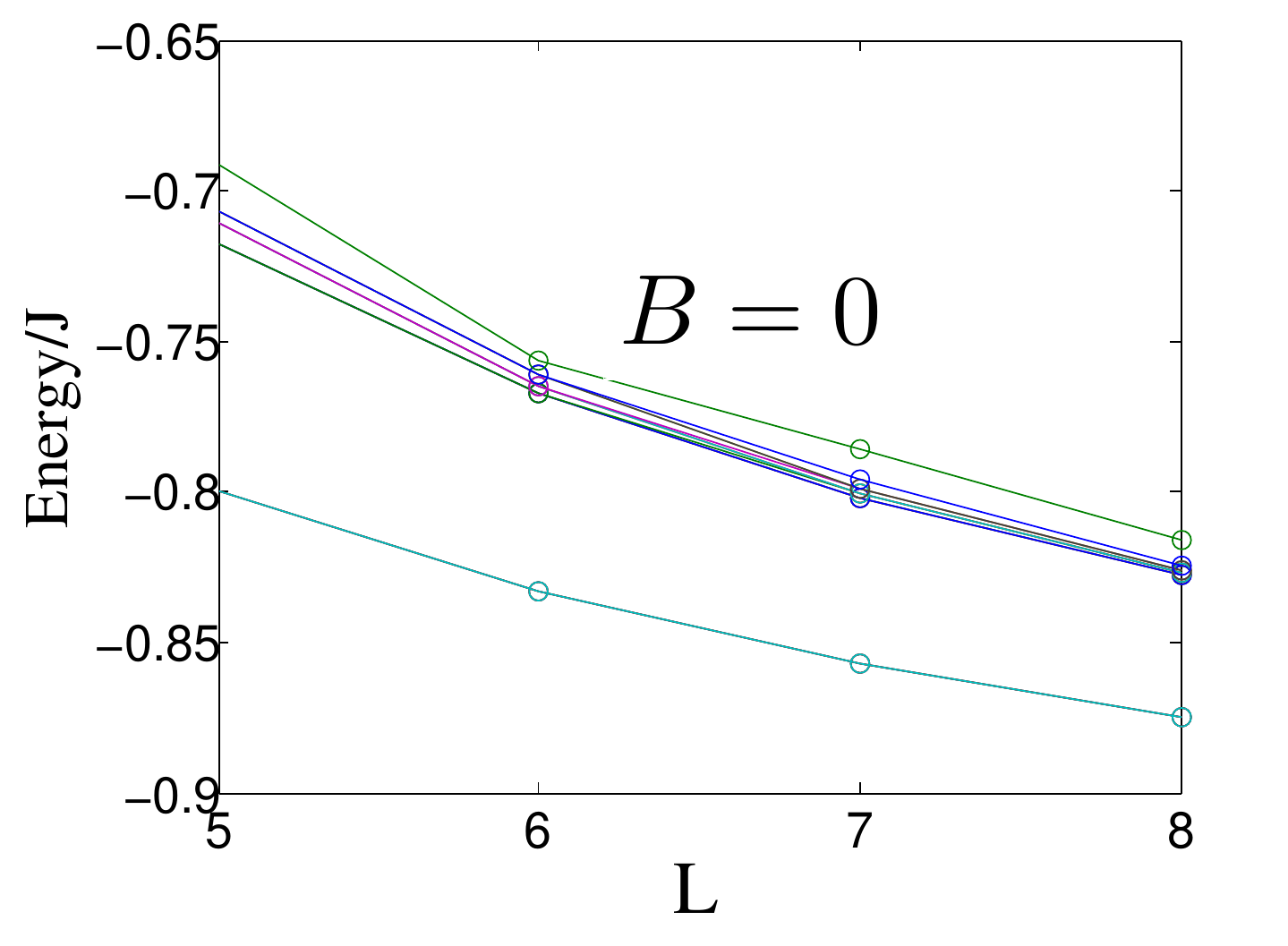}}
  \subfigure[]{
    \label{fig:class_magnetization} 
    \includegraphics[width=2.25in]{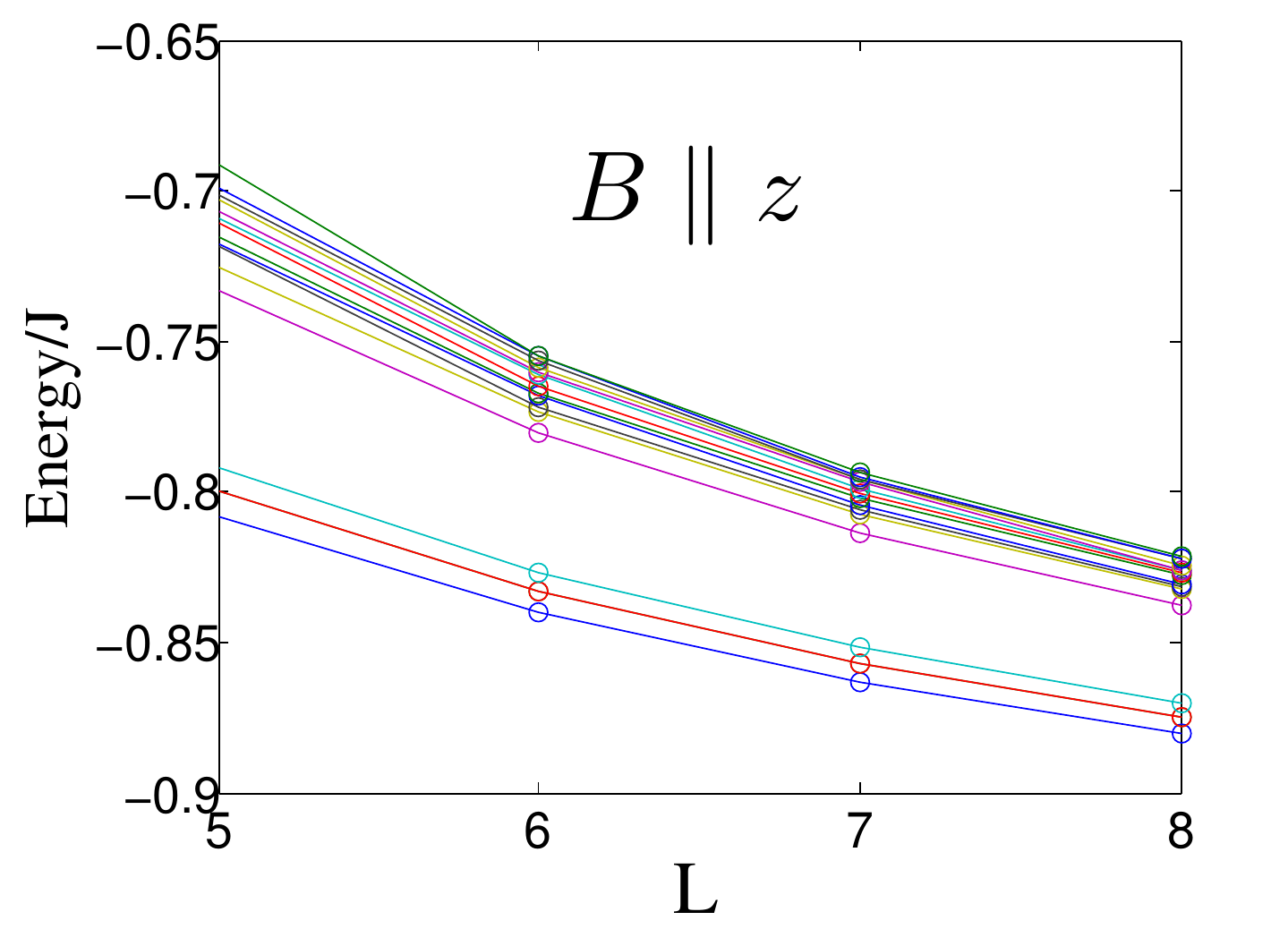}}
  \subfigure[]{
    \label{fig:class_magnetization} 
    \includegraphics[width=2.25in]{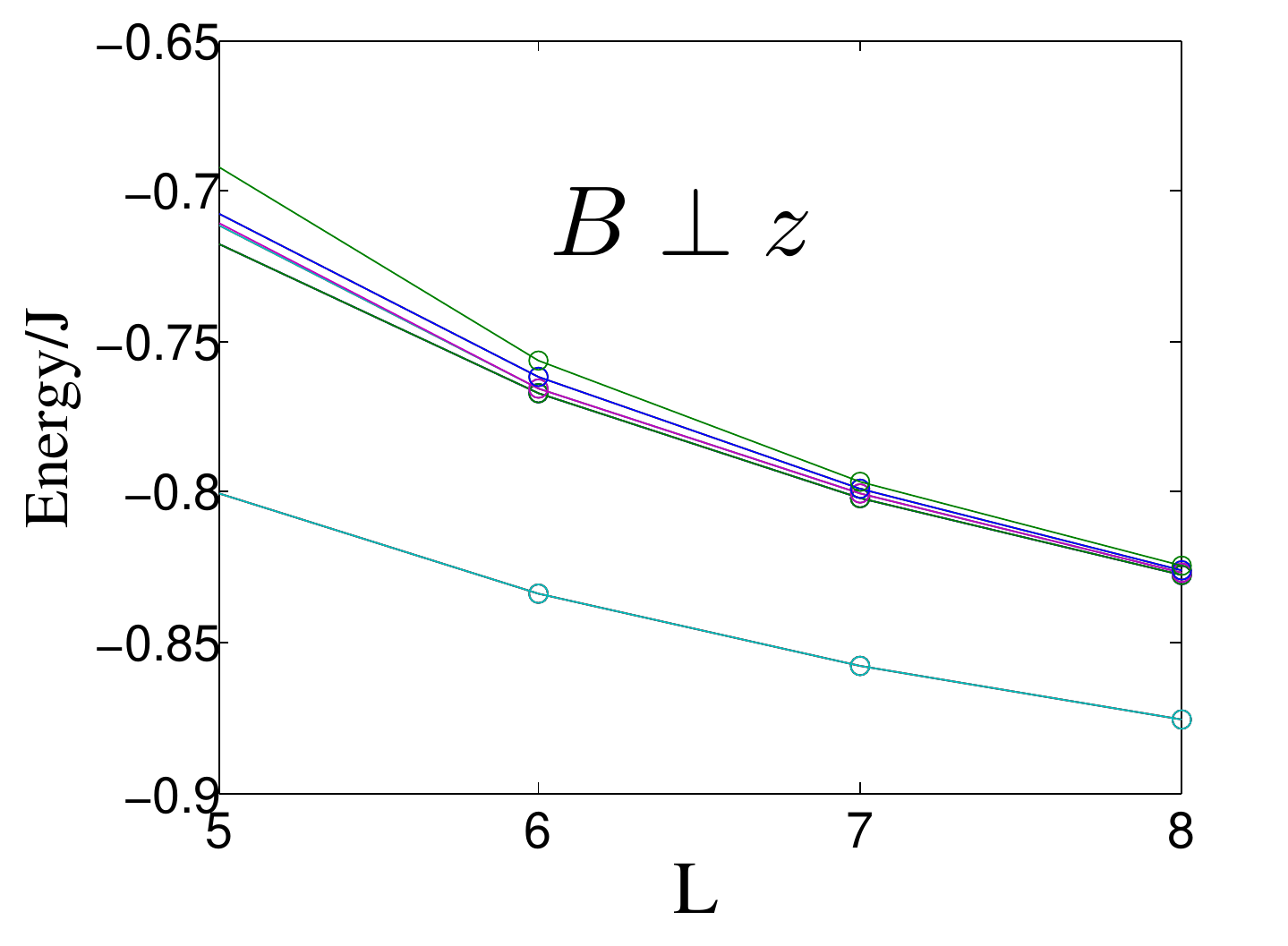}}
 \caption{The spectrum of the ground states and some excited states in $t_z$ phase, $L$ is the length of the ladder. (a) $B=0$, the ground states are 4-fold degenerate; (b) $B\parallel z$, the degeneracy of ground states is split; (c) $B\perp z$, the ground states remains degenerate. }
  \label{fig:BinTz} 
\end{figure}

This interesting result indicates that we can distinguish $t_z$ from $t_0$ by the response to magnetic fields. In $t_0$ phase, arbitrarily small magnetic field can split the degeneracy of the ground states, showing that the edge states carry free magnetic moments. According to Curie's law, the magnetic susceptibility will diverge at low temperature. But in $t_z$ phase, the edge states only carry magnetic moment in $z$ direction, so the magnetic susceptibility within the XY plane does not diverge at low temperature, but it does diverge if the magnetic field is along $z$ direction. These results are also verified numerically, see Fig.~\ref{fig:ChiTz}.
\begin{figure}[htbp]
  \centering
  \subfigure[]{
    \label{fig: link_energy} 
    \includegraphics[width=2.65in]{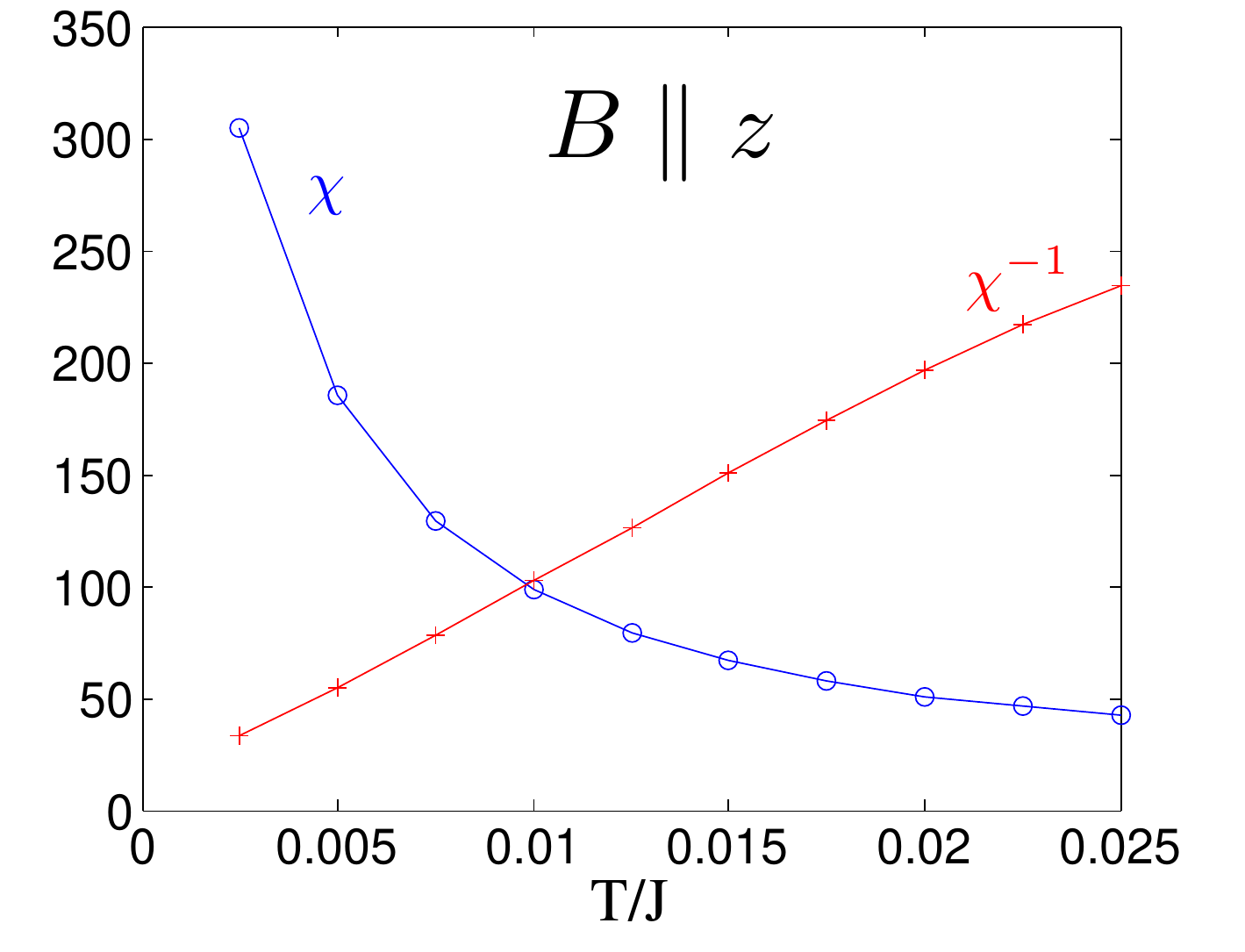}}
  \subfigure[]{
    \label{fig:class_magnetization} 
    \includegraphics[width=2.7in]{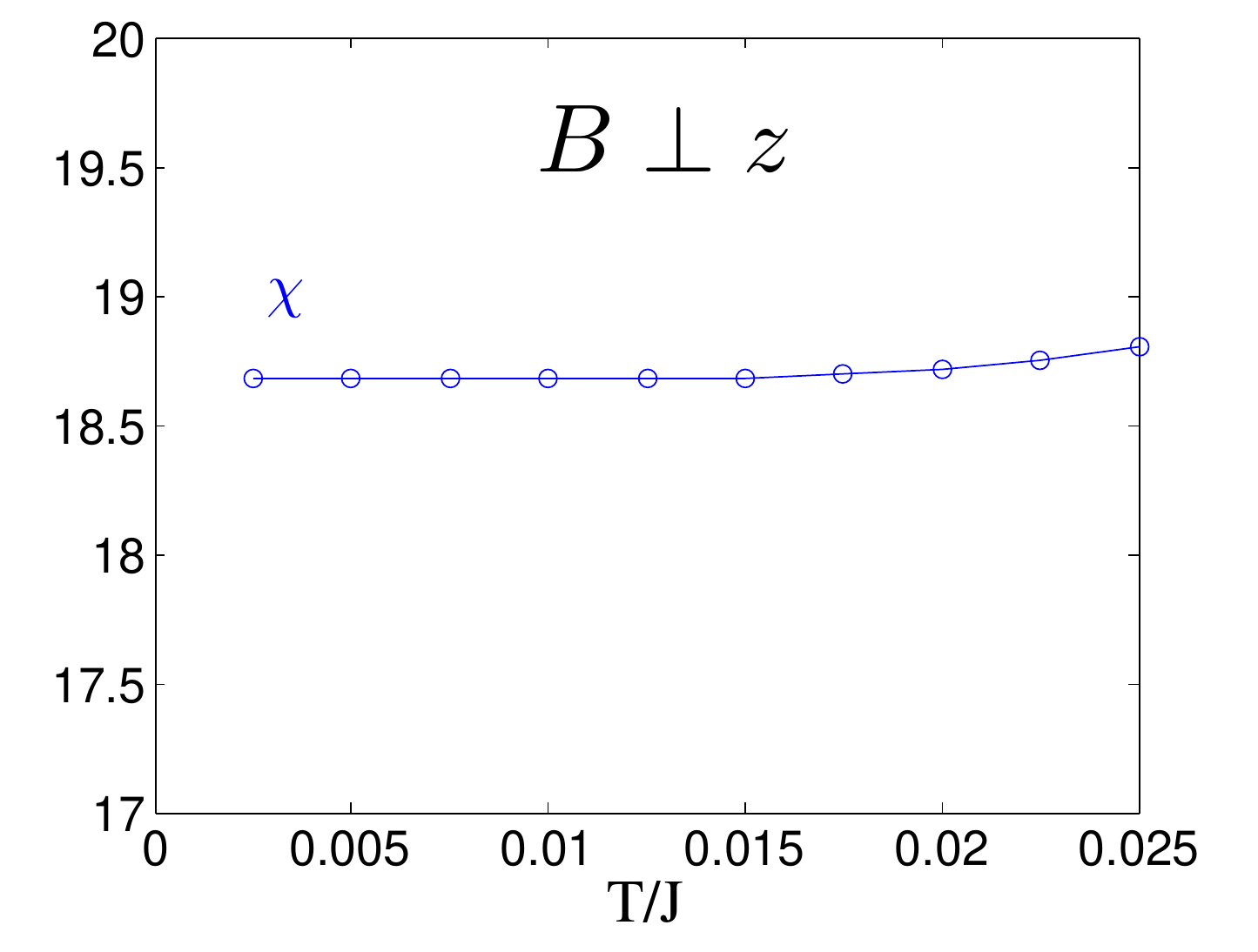}}
  \caption{The susceptibility in $t_z$ phase. (a) $\chi$ diverges at $T=0$ if $B\parallel z$; (b) $\chi$ is finite at $T=0$ if $B\perp z$.}  \label{fig:ChiTz} 
\end{figure}
The behavior of low-temperature magnetic susceptibility is measurable, which allows us to distinguish different SPT phases experimentally.
\end{widetext}

\section{Implementing the ladder Hamiltonian}

In this Appendix, we discuss in more detail the implementation scheme for the Hamiltonian with SPT phases. To keep our discussion general, we consider a two-dimensional (2D) coupled-harmonic-oscillator-array. Later, we will relate this general scheme to the specific example of a cavity-ladder as in the main text.

Consider a 2D array, on each site of the array, two independent harmonic oscillators exist and couple with those on the neighboring sites via energy tunneling. We label these harmonic oscillators as $L$ (longitudinal) and $T$ (transverse), and assume that oscillators only couple with those having the same label on the neighboring sites along the direction specified by their labels. This can be achieved by requiring the frequency difference between different types of oscillators ($L$ and $T$) to be sufficiently large, and by setting up specific coupling schemes between neighboring sites. The Hamiltonian for this 2D coupled-harmonic-oscillator-array can be written as ($\hbar $=1)
\begin{equation}
H_1=\sum_{j,k}\sum_{\mu=L,T}(v^La_{j,k}^{L+}a_{j,k+1}^L+v^Ta_{j,k}^{T+}a_{j+1,k}^T+H.c.),
\end{equation}
where $a_{j,k}^L$ ($a_{j,k}^T$) is the annihilation operator for the harmonic oscillator labeled $L$ ($T$) on the $j$th site transversally and the $k$th site longitudinally. The coupling strength along the longitudinal (transverse) direction is given by $v_L$ ($v_T$).

\begin{figure}[h]
\includegraphics[width=6cm]{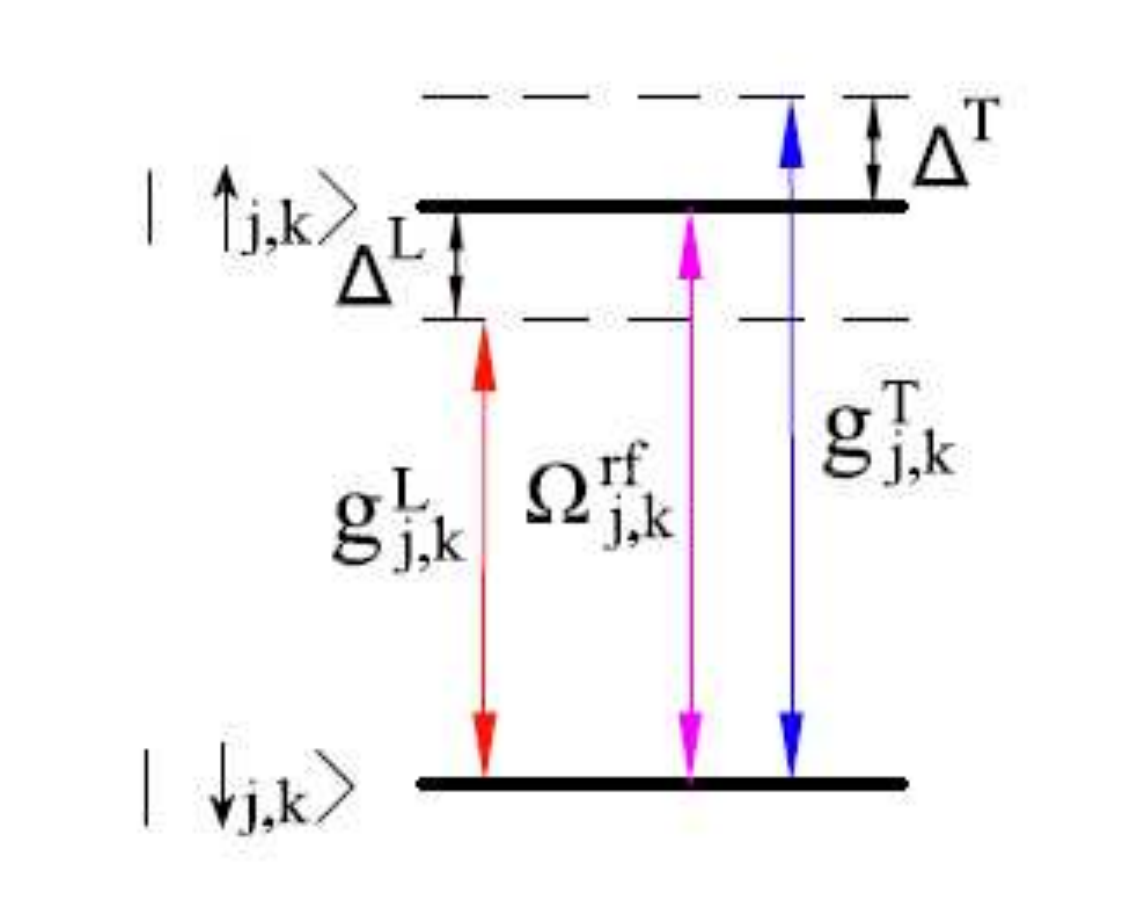}
\caption{Schematic for the coupling scheme of the two-level system on each site.}\label{twolevel}
\end{figure}

The harmonic oscillators on each site interact with a two-level system $\left\{|\uparrow_{j,k}\rangle,|\downarrow_{j,k}\rangle\right\}$, and the coupling rates are $g_{j,k}^L =g^L e^{i\theta _{j,k}^L t}$ and $g_{j,k}^T =g^Te^{i\theta _{j,k}^T t}$, respectively. This is illustrated in Fig. \ref{twolevel}. The interaction Hamiltonian is
\begin{equation}
H_2=\sum_{j,k}\sum_{\mu=L,T}(g_{j,k}^{\mu}e^{i\Delta ^{\mu}t}a_{j,k}^{\mu}S_{j,k}^{+}+H.c.),
\end{equation}
where $S_{j,k}^{+}=|\uparrow_{j,k}\rangle\langle\downarrow_{j,k}|$, and $\Delta^L$ ($\Delta^T$) denotes the detuning of the corresponding harmonic oscillator mode (see Fig. \ref{twolevel}). Finally, the two-level system on each site is coupled by a resonant dressing field, with the Hamiltonian
\begin{equation}
H_3=\sum_{j,k}\sum_{\mu=L,T}(\Omega _{j,k}^{\text{rf}}S_{j,k}^{+}+H.c.),
\end{equation}
where $\Omega _{j,k}^{\text{rf}}=\Omega^{\text{rf}} e^{i\varphi _{j,k}}$ is the Rabi frequency.

Starting from the full Hamiltonian $H=H_1+H_2+H_3$, we will eventually adiabatically eliminate the harmonic oscillator modes and derive an effective Hamiltonian for the dynamics of the coupled two-level systems throughout the array.

\begin{widetext}
Before doing so, let us first introduce the following transformations
\begin{align}
&a_{j,k}^{\mu}=\frac 1{\sqrt{MN}}\sum_{m,n}\exp [-i(\frac{2\pi jm}M+\frac{2\pi kn}N)]a_{m,n}'^{\mu}, \label{transa}\\
&\left| \uparrow _{j,k}^{\prime}\right\rangle =\frac 1{\sqrt{2}}(e^{i\varphi_{j,k} }\left| \uparrow _{j,k}\right\rangle +\left| \downarrow _{j,k}\right\rangle ),\label{transb}\\
&\left| \downarrow _{j,k}^{\prime}\right\rangle =\frac 1{\sqrt{2}}(-\left| \uparrow _{j,k}\right\rangle +e^{-i\varphi_{j,k}}\left| \downarrow _{j,k}^{\prime }\right\rangle ),\label{transc}
\end{align}
where $M$ and $N$ are the total number of sites in the longitudinal and transverse direction, respectively. While Eq. (\ref{transa}) diagonalizes $H_1$, (\ref{transb}) and (\ref{transc}) define the pseudo-spin basis $\left\{|\uparrow'_{j,k}\rangle,|\downarrow'_{j,k}\rangle\right\}$.

With these, the Hamiltonians become
\begin{align}
H_1'=&\sum_{m,n}\sum_{\mu=L,T}\omega _{m,n}^{\mu}a_{m,n}'^{\mu+}a_{m,n}'^{\mu},\\
H_2^{\prime }=& \sum_{j,k}\sum_{\mu =L,T}[\frac{g_{j,k}^\mu }{\sqrt{MN}}\sum_{m,n}\exp ^{-i(\frac{2\pi jm}M+\frac{2\pi kn}N)}a_{m,n}^{\prime \mu }e^{i\Delta ^\mu t}(e^{-i\varphi_{j,k}}\frac 1{\sqrt{2}}S_{j,k}^{\prime z}+e^{-i2\varphi_{j,k}}\frac 12S_{j,k}^{\prime +}-\frac 12S_{j,k}^{\prime -})+H.c.],\\
H_3^{\prime }=&\sum_{j,k}\sum_{\mu =L,T}\sqrt{2} \Omega^{\text{rf}} S_{j,k}^{\prime z},
\end{align}
where $\omega _{m,n}^L=2v^L\cos (\frac{2\pi n}N)$, $\omega _{m,n}^T=2v^T\cos (\frac{2\pi m}M)$. The pseudo-spin operators are defined through the pseudo-spin basis states: $S_{j,k}^{\prime +}=\left| \uparrow _{j,k}^{\prime }\right\rangle \left\langle \downarrow _{j,k}^{\prime }\right|$, and $S_{j,k}^{\prime -}=\left| \downarrow _{j,k}^{\prime }\right\rangle \left\langle \uparrow _{j,k}^{\prime }\right|$.

We now go to the rotating frame via the transformation $R=\exp [-i(H_1'+H_3')t]$
\begin{align}
H''=& R^{+}\left(\sum_iH_i'\right)R-iR^{+}\frac{dR}{dt}\nonumber\\
=&\sum_{j,k}\sum_{\mu =L,T}\{\frac{g_{j,k}^\mu }{\sqrt{MN}}\sum_{m,n}e^{-i(\frac{2\pi jm}M+\frac{2\pi kn}N)}a_{m,n}^{\prime \mu }e^{i(\Delta ^\mu -\omega _{m,n} )t}[e^{-i\varphi _{j,k}}\frac 1{\sqrt{2}}S_{j,k}^{\prime Z}+e^{-i(2\varphi _{j,k} -\sqrt{2}\Omega^{\text{rf}}  t)}\frac 12S_{j,k}^{\prime +}-\frac 12S_{j,k}^{\prime -}e^{-i\sqrt{2} \Omega^{\text{rf}}  t}]\nonumber\\
&+H.c.\}.
\end{align}

Under the condition $\left| \Delta ^T-\Delta ^L\right| \sim \left| \Delta ^\mu \right| \sim \left| \Delta ^\mu \pm \sqrt{2} \Omega^{\text{rf}}  \right| \sim \sqrt{2} \Omega^{\text{rf}}  \gg \frac{ g^\mu}{\sqrt{MN}},\omega _{m,n}^\mu $, the oscillator modes are virtually populated. We may adiabatically eliminate $a_{j,k}^{\prime \mu }$  and describe the dynamics of the system using the resultant effective Hamiltonian \cite{ZhengGuo,OsnaghiHaroche}
\begin{eqnarray}
H_{\text{eff}} &=& \sum_{j,k}\sum_{\mu =L,T}\sum_{m,n}\{\frac{(g^\mu )^2}{MN}[\frac 12\frac 1{\Delta ^\mu -\omega _{m,n}^\mu }(S_{j,k}^{\prime z})^2+\frac 14\frac 1{\Delta ^\mu -\omega _{m,n}^\mu +\sqrt{2}\Omega^{\text{rf}} }S_{j,k}^{\prime +}S_{j,k}^{\prime -}+\frac 14\frac 1{\Delta ^\mu -\omega _{m,n}^\mu -\sqrt{2}\Omega^{\text{rf}} }S_{j,k}^{\prime -}S_{j,k}^{\prime +}]\nonumber\\&&
\ +\sum_{j,k\neq j^{\prime },k^{\prime }}[\frac{\cos (\frac{2\pi (j-j^{\prime })m}M+\frac{2\pi (k-k^{\prime })n}N+\varphi _{j,k}-\varphi _{j^{\prime },k^{\prime }}+\theta _{j^{\prime },k^{\prime }}^\mu-\theta _{j,k}^{\mu})}{\Delta ^\mu -\omega _{m,n}^\mu }S_{j,k}^{\prime z}S_{j^{\prime },k^{\prime }}^{\prime z}\nonumber \\&&
\ +\frac 14(\frac{e^{-i(\frac{2\pi (j-j^{\prime })m}M+\frac{2\pi (k-k^{\prime })n}N)}e^{-i[2(\varphi _{j,k}-\varphi _{j^{\prime },k^{\prime }})+\theta _{j^{\prime },k^{\prime }}^\mu -\theta _{j,k}^\mu ]}}{\Delta ^\mu -\omega _{m,n}^\mu +\sqrt{2}\Omega^{\text{rf}} }\ +\frac{e^{i[\frac{2\pi (j-j^{\prime })m}M+\frac{2\pi (k-k^{\prime })n}N]}e^{i(\theta _{j^{\prime },k^{\prime }}^\mu -\theta _{j,k}^\mu )}}{\Delta ^\mu -\omega _{m,n}^\mu -\sqrt{2}\Omega^{\text{rf}} })S_{j,k}^{\prime +}S_{j^{\prime },k^{\prime }}^{\prime -}\nonumber \\&&
+\frac 14(\frac{e^{-i(\frac{2\pi (j-j^{\prime })m}M+\frac{2\pi (k-k^{\prime })n}N)}e^{-i(\theta _{j^{\prime },k^{\prime }}^\mu -\theta _{j,k}^\mu )}}{\Delta ^\mu -\omega _{m,n}^\mu -\sqrt{2}\Omega^{\text{rf}} }\ \ +\frac{e^{i(\frac{2\pi (j-j^{\prime })m}M+\frac{2\pi (k-k^{\prime })n}N)}e^{i[2(\varphi _{j,k}-\varphi _{j^{\prime },k^{\prime }})+\theta _{j^{\prime },k^{\prime }}^\mu -\theta _{j,k}^\mu ]}}{\Delta ^\mu -\omega _{m,n}^\mu +\sqrt{2}\Omega^{\text{rf}} })S_{j,k}^{\prime -}S_{j^{\prime },k^{\prime }}^{\prime +}]\}.\nonumber\\&&
\end{eqnarray}
Adopting the formulae: $\sum_n\cos (\frac{2\pi n}N)\equiv 0$, $\sum_n\cos ^2(\frac{2\pi n}N)=\sum_n\cos (\frac{2\pi n}N)e^{\pm i\frac{2\pi n}N}\equiv \frac N2$, and keeping only on-site and nearest-neighbor interactions, we can further simplify the effective Hamiltonian
\begin{eqnarray}
H_{\text{eff}} &\simeq& \sum_{j,k}\sum_{\mu =L,T}\{\frac{ (g^\mu)^2\Delta ^\mu }{2[(\Delta ^\mu )^2-2 (\Omega^{\text{rf}}) ^2]}(S_{j,k}^{^{\prime }})^2-\frac{(g^\mu)^2 (\Omega^{\text{rf}})^2}{\Delta ^\mu [(\Delta ^\mu )^2-2(\Omega^{\text{rf}})^2]}(S_{j,k}^{\prime z})^2-\frac{(g^\mu)^2 \Omega^{\text{rf}}  }{2\sqrt{2}[(\Delta ^\mu )^2-2(\Omega^{\text{rf}})^2]}S_{j,k}^{^{\prime }z}\}\nonumber\\&&
-(-1)^{\frac{\left| \theta _{j,k+1}^L-\theta _{j,k}^L\right| }\pi }\sum_{j,k}(g^L)^2\{(-1)^{\frac{\left| \varphi _{j,k+1}-\varphi _{j,k}\right| }\pi }\frac{v^L}{(\Delta ^L)^2}S_{j,k}^{\prime z}S_{j,k+1}^{\prime z}+\frac{v^L[(\Delta ^L)^2+2(\Omega^{\text{rf}}) ^2]}{[(\Delta ^L)^2-2(\Omega^{\text{rf}}) ^2]^2}(S_{j,k}^{\prime x}S_{j,k+1}^{\prime x}+S_{j,k}^{\prime y}S_{j,k+1}^{\prime y})\}\nonumber\\&&
-(-1)^{\frac{\left| \theta _{j+1,k}^T-\theta _{j,k}^T\right| }\pi }\sum_{j,k}(g^T)^2\{(-1)^{\frac{\left| \varphi _{j+1,k}-\varphi _{j,k}\right| }\pi }\frac{v^T}{(\Delta ^T)^2}S_{j,k}^{\prime z}S_{j+1,k}^{\prime z}+\frac{v^T[(\Delta ^T)^2+2(\Omega^{\text{rf}}) ^2]}{[(\Delta ^T)^2-2(\Omega^{\text{rf}}) ^2]^2}(S_{j,k}^{\prime x}S_{j+1,k}^{\prime x}+S_{j,k}^{\prime y}S_{j+1,k}^{\prime y})\}.\nonumber\\&&\label{effHsupp}
\end{eqnarray}
While the first two terms on the first line in Eq. (\ref{effHsupp}) are constant and can be dropped, the third term is a Stark-shift in the pseudo-spin basis, and can be canceled via local optical elimination.~\cite{ChoBose} This corresponds to applying a r.f. or Raman fields with appropriate magnitude and phase between the hyperfine states such that the effective Stark-shift is canceled.  Then, under the condition $3(\Delta ^\mu )^2=2(\Omega^{\text{rf}})^2$, we have
\begin{eqnarray}
H_{\text{eff}} &\simeq& -(-1)^{\frac{\left| \theta _{j,k+1}^L-\theta _{j,k}^L\right| }\pi }\frac{3(g^L)^2v^L }{2(\Omega^{\text{rf}})^2}\sum_{j,k}[(-1)^{\frac{\left| \varphi _{j,k+1}-\varphi _{j,k}\right| }\pi }S_{j,k}^{\prime z}S_{j,k+1}^{\prime z}+S_{j,k}^{\prime x}S_{j,k+1}^{\prime x}+S_{j,k}^{\prime y}S_{j,k+1}^{\prime y})]\nonumber\\&&
-(-1)^{\frac{\left| \theta _{j+1,k}^T-\theta _{j,k}^T\right| }\pi }\frac{3(g^T)^2v^T }{2(\Omega^{\text{rf}})^2} \sum_{j,k}[(-1)^{\frac{\left| \varphi _{j+1,k}-\varphi _{j,k}\right| }\pi }S_{j,k}^{\prime z}S_{j+1,k}^{\prime z}+S_{j,k}^{\prime x}S_{j+1,k}^{\prime x}+S_{j,k}^{\prime y}S_{j+1,k}^{\prime y}].\label{effHsupp2}
\end{eqnarray}

Eq. (\ref{effHsupp2}) gives the most general form of the effective Hamiltonian using our setup. For the spin-ladder Hamiltonians we considered in the main text, we may take $M=2$ so that $j=1,2$ in the summations. Then, depending on the magnitudes and the relative phases of the coupling fields, we have either the Hamiltonian for the $t_0$ phase ($t_0$ model) or the Hamiltonian for the $t_z$ phase ($t_z$ model). In particular, with $\left| \theta _{j,k+1}^L -\theta _{j,k}^L \right|=\pi $, $\theta_{j,k}^T=\varphi_{j,k}=0$ for arbitrary $\{j,k\}$, the Hamiltonian reduces to the $t_0$ model
\begin{eqnarray}
H_0 &=& J\sum_{j=1}^2\sum_k(S_{j,k}^{\prime z}S_{j,k+1}^{\prime z}+S_{j,k}^{\prime x}S_{j,k+1}^{\prime x}+S_{j,k}^{\prime y}S_{j,k+1}^{\prime y})\nonumber\\&&
+\lambda \sum_k(S_{1,k}^{\prime z}S_{2,k}^{\prime z}+S_{1,k}^{\prime x}S_{2,k}^{\prime x}+S_{1,k}^{\prime y}S_{2,k}^{\prime y}),
\end{eqnarray}
where the interaction rate $J=\frac{3(g^L)^2v^L}{2(\Omega^{\text{rf}})^2}$ and $\lambda =-\frac{3(g^T)^2v^T}{2(\Omega^{\text{rf}})^2}$.  On the other hand, when $\left| \theta _{j,k+1}^L -\theta _{j,k}^L \right|=\left|\theta_{1,k}^{T}-\theta_{2,k}^T\right| =\pi$, $\left| \varphi _{1,k}-\varphi _{2,k}\right| =\pi $ and $\varphi _{j,k}=\varphi _{j,k+1}$, the Hamiltonian reduces to the $t_z$ model
\begin{eqnarray}
H_z &=& J\sum_{j=1}^2\sum_k(S_{j,k}^{\prime z}S_{j,k+1}^{\prime z}+S_{j,k}^{\prime x}S_{j,k+1}^{\prime x}+S_{j,k}^{\prime y}S_{j,k+1}^{\prime y})\nonumber\\&&
+\lambda \sum_k(S_{1,k}^{\prime z}S_{2,k}^{\prime z}-S_{1,k}^{\prime x}S_{2,k}^{\prime x}-S_{1,k}^{\prime y}S_{2,k}^{\prime y}).
\end{eqnarray}
\end{widetext}
A straightforward example for the realization of the coupled-harmonic-oscillator-array is the coupled quantum electrodynamics (QED) cavity array, as shown in FIG. 5(a) in the main text. In such a system, atoms or solid spins interact with the quantized cavity fields, which couple to their neighboring ones across both the longitudinal and transverse directions via photon hopping.~\cite{HartmannPlenio} The two-level system in our general model can be replaced by a three-level structure, with two low-lying hyperfine states and an electronically excited state. Correspondingly, the coupling $g^{L}_{j,k}$ ($g^{T}_{j,k}$) in the general model is replaced by a Raman path in the longitudinal (transverse) direction, with an external laser field and a cavity mode each contributing a leg in the Raman coupling. Note that due to the large difference in the two-photon detuning of the Raman couplings, the two Raman paths are effectively independent. It is then straightforward to work out the correspondence: $g^\mu =\frac{G^\mu \Omega ^\mu }2(\frac 1{\Delta _1^\mu }+\frac 1{\Delta _2^\mu })$, $\left|\Delta^{\mu}\right|=\left|\Delta^{\mu}_2-\Delta^{\mu}_1\right|$, where $G^{\mu}$ is the Rabi frequency for the atom-cavity coupling, $\Delta^{\mu}$ is the detuning (c.f. Fig. 5 in the main text). Importantly, one may realize Hamiltonians of different SPT phases ($t_0$ or $t_z$) by adjusting the phases of the Rabi frequencies $\Omega^{\mu}_{j,k}$ and $\Omega^{\text{rf}}_{j,k}$. For typical experimental parameters \cite{BoozerKimble}: $\Omega ^\mu\sim 100$MHz, $G^{\mu}\sim 100$MHz, $\left|\Delta^{\mu}_i\right|\sim 1$GHz ($i=1,2$), $v^{\mu}\sim10$MHz, $\Omega ^{\mathrm{rf}} \sim 100$MHz, we have $J\sim 0.15$MHz, with the magnitude of $\lambda/J$ widely tunable by adjusting the ratio between $\Omega^L$ and $\Omega^T$.




\begin{thebibliography}{99}
\bibitem{note2}The anisotropic Heisenberg exchange interactions $J_x S^x_iS^x_j+J_y
S^y_iS^y_j+J_z S^z_iS^z_j$  considered in this paper actually have a larger
symmetry group, $D_2\times\sigma\times T$.  The full symmetry group
$D_2\times\sigma\times T$ has 128 different projective representations that
correspond to one trivial phase and 127 non-trivial SPT phases. All of these
phases can be realized in spin ladders, but only part of them can be realized in
two-legged ladders. The $t_0, t_x, t_y, t_z$ phases discussed in this paper
belong to the 127 non-trivial phases.
\bibitem{note1} The three phases $t_x, t_y, t_z$ have similar properties with the $T_x, T_y, T_z$ phases, respectively, discussed in Ref.~\onlinecite{LCW} for spin-1 chain models.
\end{thebibliography}
\end{document}